\documentclass[final,onefignum,onetabnum]{siamonline171218}

\usepackage{amsmath,bm}

\usepackage{amsfonts}
\usepackage{epstopdf}
\usepackage{graphicx}
\usepackage{url}
\usepackage{dsfont}
\usepackage{xcolor}
\usepackage{subcaption}
\usepackage{setspace}
\usepackage{booktabs}
\DeclareMathAlphabet\mathpzc{OT1}{pzc}{m}{it}
\let\mathcal=\mathpzc
\def\E{{\mathbb E}}

\let\trueiiint=\iiint
\def\iiint{\mathop{\textstyle\trueiiint}\limits}
\def\intinfty{\int\limits_{\!\!-\infty\,\,}^{\,\,\infty\!\!}\kern-0.0em}
\def\iintinfty{\mathop{\int\!\!\int}\limits_{\!\!-\infty\,\,}^{\,\,\infty\!\!}\kern-0.0em}
\def\iiintinfty{\mathop{\int\!\!\int\!\!\int}\limits_{\!\!-\infty\,\,}^{\,\,\infty\!\!}\kern-0.0em}

\let\<=\langle
\let\>=\rangle
\let\^=\hat
\def\~#1{{\-ox{\sf#1}}}

\let\-=\mathbf

\def\bolde{{\boldsymbol\epsilon}}

\def\circ{\ifmmode\mathchar"220E\else$\mathchar"220E$\fi}

\def\@#1{{\bf #1}}

\usepackage{lipsum}
\usepackage{amsfonts}
\usepackage{graphicx}
\usepackage{epstopdf}

\usepackage[algo2e,linesnumbered,boxed,ruled,commentsnumbered]{algorithm2e}
\usepackage{algpseudocode,float}
\ifpdf
  \DeclareGraphicsExtensions{.eps,.pdf,.png,.jpg}
\else
  \DeclareGraphicsExtensions{.eps}
\fi

\usepackage{enumitem}
\setlist[enumerate]{leftmargin=.5in}
\setlist[itemize]{leftmargin=.5in}

\newsiamremark{remark}{Remark}
\newsiamremark{hypothesis}{Hypothesis}
\crefname{hypothesis}{Hypothesis}{Hypotheses}
\newsiamthm{claim}{Claim}

\headers{A defensive marginal particle filter}{L. Wen, J. Wu, L. Lu and J. Li}

\title{A defensive marginal particle filtering method for data assimilation\thanks{Submitted to the editors DATE.
\funding{This work was supported by the NSFC under grant number 11301337 and 5150080805.}}}

\author{Linjie Wen\thanks{School of Mathematical Sciences and Institute of Natural Sciences, 
Shanghai Jiao Tong University, 800 Dongchuan Rd, Shanghai 200240, China. }
\and Jiangqi Wu\footnotemark[2]
\and Linjun Lu\thanks{Corresponding author, School of Naval Architecture, Ocean and Civil Engineering, Shanghai Jiao Tong University, Shanghai 200240, China 
\email{linjunlu@sjtu.edu.cn}.}
\and Jinglai Li\thanks{Corresponding author, Department of Mathematical Sciences, University of Liverpool, Liverpool, UK  
\email{jinglai.li@liverpool.ac.uk}.}}
\usepackage{amsopn}

\ifpdf
\hypersetup{
  pdftitle={A defensive marginal particle filter},
  pdfauthor={L. Wen, J. Wu, L. Lu and J. Li}
}
\fi

\begin{document}

\maketitle

\begin{abstract}
Particle filtering (PF) is an often used method to estimate the states of dynamical systems. 
A major limitation of the standard PF method is that the dimensionality of the state space increases as the time proceeds and 
eventually may cause degeneracy of the algorithm. 
A possible approach to alleviate the degeneracy issue is to compute the marginal posterior distribution at each time step, which leads to the so-called marginal PF method.
A key issue in the marginal PF method is to construct a good sampling distribution in the marginal space.
When the posterior distribution is close to Gaussian, the Ensemble Kalman filter (EnKF) method can usually provide a good sampling distribution; however 
the EnKF approximation may fail completely when the posterior is strongly non-Gaussian.  
In this work we propose a defensive marginal PF (DMPF) algorithm which constructs a sampling distribution in the marginal space by combining the standard PF and the EnKF approximation using a multiple importance sampling (MIS) scheme.
An important feature of the proposed algorithm is that it can automatically adjust the relative weight of the PF and the EnKF components 
in the MIS scheme in each step, according to how non-Gaussian the posterior is.  
With numerical examples we demonstrate that the proposed method can perform well regardless of whether the posteriors can be well approximated by Gaussian. 
\end{abstract}



\section{Introduction} \label{section_intro}
Assimilation of data into mathematical models is an essential task in many fields of science and engineering, ranging from meteorology~\cite{ghil1991data} to robotics~\cite{thrun}. 
Simply speaking, data assimilation is to estimate the optimal prediction based on both the output of the mathematical model, which is  only an approximation of the real-world system, and the observations that are subject to measurement noise~\cite{law2015data}.
The Kalman filter(KF) type of methods, which are based on linear control theory and optimization are a popular tool for data assimilation problems. Unfortunately, it is usually challenging to apply such methods to nonlinear systems, as they often require some linearization or approximation processes, e.g. the extended Kalman filter~\cite{jazwinski} or the ensemble Kalman filter~\cite{evensen2009data}. Sometimes these methods can even fail~\cite{kuzn,salman} when strong nonlinearity is present.

On the other hand, the sequential Monte Carlo (SMC) method~(see e.g., \cite{SMC}), also known as particle filtering (PF), can deal with problems with strongly nonlinear models, without any linearization or approximation. The basic idea of PF is the following.
Suppose that the mathematical model is a nonlinear stochastic dynamical system (more details about the mathematical model are provided
in Section~\ref{sec:ssmodel} with the specific formulation of the model given by Eqs.~\eqref{e:state}),
and our goal is to estimate the hidden states $\{\-u_t\}_{t=0}^T$ of the system from noisy partial observations $\{\-y_t\}_{t=0}^T$ of the system. 
This can be done with the so-called Bayes filter (also known as the optimal filter), where
the posterior probability density function (PDF) of the hidden states is estimated 
by the Bayes' rule recursively~\cite{doucet}. 
As the posterior distribution usually does not admit an analytical form,
the PF method approximates the posterior distribution with Monte Carlo sampling~(hence its name SMC). 
That is, PF employs a number of independent random realizations
called particles, sampled directly from the state space, to represent
the posterior probability: namely, at each time $t$, the method first generates particles and then updates the weight of each particle according to the 
observations $\-y_t$. 
 For further discussions on the PF method and its applications, 
we refer to \cite{SMC,arulampalam2002tutorial,doucet2009tutorial,cappe2007overview} and
the references therein.

The PF method in its very basic form can be understood as {\color{black}drawing} weighted samples according to the joint distribution $\pi(\-u_{0:T}|\-y_{0:T})$ using the importance sampling (IS) technique.
When $T$ is large, the method thus performs IS simulations in a high-dimensional state space, which may result in degeneracy of the particles (the IS weights becoming zero for all but one particle)~\cite{doucet2009tutorial}.  
On the other hand, often in practice one is only interested in the marginal distribution $\pi(\-u_t|\-y_{0:t})$, which implies that it is unnecessary to sample the high-dimensional joint distribution $\pi(\-u_{0:T}|\-y_{0:T})$. 
Instead, one can perform IS only in the marginal space of {{$\pi(\-u_t|\-y_{0:t})$}}, and based on this idea, a method called marginal particle filter (MPF) was proposed in \cite{klaas2005toward} to alleviate the degeneracy issue. 
The method later has found applications in the estimation of filter derivative~\cite{poyiadjis2005particle} and robot localization~\cite{martinez2007analysis}.
A key in the MPF method is to construct an IS or proposal distribution that can approximate well the marginal posterior $\pi(\-u_t|\-y_{0:t})$
 at each time step. 
One very natural idea is to construct the proposal distribution using the ensemble Kalman 
filter~(EnKF). The basic idea behind EnKF is to assume the posterior distribution at each step follows a Gaussian distribution with the mean and covariance estimated from samples, and then update the samples according to the Kalman filter formulation.
{{\color{black} We emphasize here that, the difference between a direct use of EnKF and 
using it as an proposal in a PF/SMC scheme is that the SMC scheme can correct for the bias in 
the EnKF particles (due to the Gaussian approximation) 
by assigning an IS weight to each particle. 
The idea of using the EnKF approximation as a proposal distribution in PF is not new: for example, it has been used 
in \cite{papadakis2010data} to construct an independent sequential IS distribution in the PF framework,
and later its use in the MPF scheme is discussed in \cite{morzfeld2017collapse}. 
 A limitation of the EnKF based proposal distribution is that (just like the EnKF method itself) it may result in extremely poor estimates or even fail completely (see the example in Section~\ref{sec:bern}) when the posterior is strongly non-Gaussian;
unfortunately it is almost impossible to {\color{black}know} whether the posteriors are {\color{black}close} to Gaussian in advance.
The main contribution of the work is to propose defensive scheme to prevent such a failure of the estimation:
it can automatically adjust between a proposal based on the EnKF approximation and a standard PF proposal,
and as a result, the proposed algorithm may perform well regardless of whether the posteriors 
are close to Gaussian.}
Specifically  our defensive scheme combines the EnKF based proposal and the standard PF proposal using the multiple IS method (also known 
as the deterministic mixture). 
In fact, combing several different IS distributions using multiple IS~\cite{veach1995optimally} to prevent 
the risk of failure of a single IS distribution is a very popular safeguard measure in the IS literature, e.g., \cite{hesterberg1995weighted,owen2000safe,cornuet2012adaptive},
and the use of MPF here makes it possible to implement the method in the filtering problems. 
A key issue in the multiple IS method is to determine the weight of each component (in the present setting 
the two components are EnKF and PF respectively).  
Ideally at each time step we want the method to choose the EnKF approximation if the posterior distribution is close to Gaussian and the standard PF otherwise.  
To achieve this goal, we here provide an algorithm that can automatically determine the relative weight between the EnKF and the PF components by minimizing the variance of the resulting IS weight. 
Finally it is important to note that the proposed defensive MPF scheme does not depend on the EnKF approximation,
and it can be used with any proposal distribution constructed in the marginal space. 

The rest of the paper is arranged as follows. In Section~\ref{Sec:pf}, we first introduce the basic setup of the filtering problem of dynamical models and then
discuss the standard PF and EnKF methods for solving this type of problems.
In Section~\ref{Sec:mpf}, we present in detail our defensive MPF method.
Numerical examples are provided in Section \ref{Sec:examples} 
to compare the performance of the proposed method and the existing ones,
and finally Section~\ref{Sec:conclusion} offers some concluding remarks.

\section{The PF and the EnKF methods}\label{Sec:pf}

We give a brief overview of the formulation of the PF and the EnKF methods in this section.
   
\subsection{The problem seteup}\label{sec:ssmodel}
We consider the  filtering problem in 
a generic form:
\begin{subequations}
\label{e:state}
\begin{align}
&{\color{black} \-u_t \sim f_t(\cdot|\textbf{u}_{t-1}), \quad  \-u_0\sim \pi_0(\-u_0)},\label{cmodel}\\
&\-y_{t}=H_{t}\textbf{u}_{t} + \bm{\eta}_t\label{cmeasure},
\end{align}
\end{subequations}
where $\textbf{u}_t\in \mathbb{R}^{n_{u}}$ denotes the state vector at time $t$,
$\-y_t\in \mathbb{R}^{n_{v}}$ 
is the observed data at time $t$, 
$f_t(\cdot|\textbf{u}_{t-1})$ is the distribution of $\-u_t$ conditional on $\-u_{t-1}$, 
 $H_t$, a {\color{black}${n_v}\times {n_u}$} matrix, is the observation operator at time $t$,
and $\bm{\eta}_t$ is the observation noise.
{\color{black}In this work, we shall assume that the observation noise $\bm{\eta}_t$ is Gaussian and
 the noise at different time steps is independent from each other. }
In a filtering problem, the observation $\-y_t$ arrives sequentially in time
and the goal is to estimate the true state $\textbf{u}_t$, based
on the prediction by \eqref{cmodel} and the measurement \eqref{cmeasure}.
Finally we emphasize here that the dynamic model \eqref{cmodel} is Markovian, in that any future $\textbf{u}_{t+1}$ is independent of the past given the present $\textbf{u}_{t}$, and the observation $\-y_{t+1}$ is independent from $\-y_{0:t}$ conditional on $\textbf{u}_{t}$:
\begin{equation}
\pi(\textbf{u}_{t+1} | \textbf{u}_{0:t}, \-y_{0:t}) = \pi(\textbf{u}_{t+1}|\textbf{u}_{t}), \,\, \mbox{and}\,\, \pi(\-y_{t+1}|\-u_{0:t+1},\-y_{0:t}) = {\color{black}\pi(\-y_{t+1}|\-u_{t+1})}, 
\label{e:markov}
\end{equation}
which will be used in the derivation of the PF method. 
{\color{black} Note here that throughout this paper we use $\pi$ as a generic notation of 
probabilistic distribution,  the actual meaning of which is specified by its arguments.}

\subsection{The particle filter} \label{sec:smc}
In general, we can formulate the filtering problem in a Bayesian inference framework: i.e., we try to infer state parameters $\-u_{0:T}$ from data $\-y_{0:T}$ for some positive integer $T$, and ideally we can compute the posterior distribution using the Bayes' formula: 
\[\pi(\-u_{0:T}|\-y_{0:T})  = \frac{\pi(\-y_{0:T}|\-u_{0:T}) \pi({\color{black}\-u}_{0:T})}{\pi(\-y_{0:T})}.
\]
The PF (or SMC) method allows to generate (weighted) samples, called particles, from the posterior distribution 
$\pi(\-u_{0:T}|\-y_{0:T})$, which can be used to evaluate any quantities of interest associated with the 
posterior $\pi(\-u_{0:T}|\-y_{0:T})$.

We now give a brief overview of the PF method, and it is easier to start with a standard MC estimation. 
Suppose that there is a real-valued function $h(\cdot): \-R^{T\times n_u}\rightarrow R$ and  we are interested in the expectation 
\[ I = \E_{u_{0:T}|y_{0:T}}[h(\-u_{0:T})] = \int h(\-u_{0:T}) \pi(\-u_{0:T}|\-y_{0:T}) d\-u_{0:T}\]
which can be estimated with a MC estimator: 
\[ \hat{I} =  \frac1M\sum_{m=1}^M h(\-u_{0:T}^m), 
\]
where $\{\-u_{0:T}^m\}_{m=1}^M$ are samples drawn from $\pi(\-u_{0:T}|\-y_{0:T})$. It should be clear that the MC estimator $\hat{I}$ is an unbiased estimator of $I$. 
In many practical problems, drawing samples from the target distribution $\pi(\-u_{0:T}|\-y_{0:T})$ can be a challenging task, and in this case, we can use the technique of {\color{black}IS}. The  IS method introduces an importance distribution $q_t(\-u_{0:T})$ and rewrites
\[
I  = \int h(\-u_{0:T}) \pi(\-u_{0:T}|\-y_{0:T}) d\-u_n = \int h(\-u_{0:T}) w(\-u_{0:T}) q(\-u_t|\-y_{0:T}) d\-u_t \]
with $w_t(\-u_{0:T}) = \pi(\-u_{0:T}|\-y_{0:T})/q_t(\-u_{0:T})$ is the IS weight. It yields directly an IS estimator of $I$: 
\[\hat{I}_{IS} =  \frac1M\sum_{m=1}^M h(\-u_{0:T}^m) w(\-u_{0:T}^m), \]
where the samples{\color{black} $\{\-u_{0:T}^m\}_{m=1}^M$} are drawn from the importance distribution $q_t(\-u_{0:T})$,
and it can also be verified that the IS estimator is also an unbiased one for $I$. 
The IS requires to generate samples from $q(\-u_{0:T})$ and  to draw the joint sample $\-u_{0:T}$ from a joint distribution~$q(\-u_{0:T})$. 
Using the Markovian property in Eq.~\eqref{e:markov}, we can write the posterior distribution $\pi(\-u_{0:T}|\-y_{0:T})$  in the form of 
\begin{equation}
\pi(\-u_{0:T}|\-y_{0:T}) = \frac1{Z_T}
 \pi(\-y_0|\-u_{0}) \pi(\-u_0)\prod_{t=1}^{T} {\pi(\-y_{t}|\-u_{t}) \pi(\-u_{t}|\-u_{t-1})}, 
 \label{e:jointpost}
 \end{equation}
 where $Z_T$ is the normalization constant. 
Similarly, we can also assume that the importance distribution 
$q(\-u_{0:T})$ is also given in such a sequential  form:
\[q(\-u_{0:T}) = q_0(\-u_0)\prod_{t=1}^{T} q_{t}(\-u_{t}|\-u_{t-1}), \]
and the resulting IS weight function is
\begin{equation}
w_0(\-u_0) =\frac{\pi(\-y_0|\-u_0) \pi(\-u_0)}{q_0(\-u_0)},\label{e:w0}
\end{equation}
and 
\begin{equation}
w_T(\-u_{0:T}) = \frac{1}{Z_T}w_0(\-u_{0}) \prod_{t=1}^{T}\alpha_{t}(\-u_{0:t}),\label{e:wupdate}
\end{equation}
for $t>0$, 
where  $\alpha_{t}$ is the incremental weight function:
\begin{equation}
 \alpha_{t} (\-u_{0:t}) = \frac{\pi(\-y_t|\-u_t) \pi(\-u_t|\-u_{t-1})}{q_t(\-u_{t}|\-u_{t-1})}. \label{e:alpha}
 \end{equation}
   We note that, in the formulation above, we do not have the knowledge of the normalization constant $Z_T$.  
 In the implementation, however, we can simply set the normalization constant to be 1, and renormalize
 the weights computed. Namely, suppose that we draw a set of samples $\{\-u_{0:T}^m\}_{m=1}^M$ from the IS distribution $q(\-u_{0:T})$,  
 and we compute the weights $\{w_{T}^m\}_{m=1}^M$ of  the samples using Eqs.~\eqref{e:w0}-\eqref{e:alpha} (and taking $Z_T=1$), and then renormalize the weights as,  
 \begin{equation}
 {\color{black}W_T^m = \frac{w_T^m}{\sum_{m'=1}^M w_T^{m'}}.}
 \end{equation}
%
The PF algorithm performs the procedure described above in a recursive manner: \\
{{  
\begin{algorithm}[H]
\caption{The PF algorithm}
\begin{enumerate}
\item At $t=0$, sample $\{\textbf{u}^m_{0}\}_{m=0}^{M} \sim q_0(\textbf{u}_0)$, and compute $\{w_0^m= w_0({\color{black}\textbf{u}}_0^m)\}_{m=1}^M$ using Eq~\eqref{e:w0}; renormalize the weights: $W_0^m = \frac1{\sum_{m'=1}^M w_0^{m'}} w_0^m$.

\item At $t>1$:

\item  prediction step: for each $m=1...M$, draw $\textbf{u}_t^m \sim q_t(\textbf{u}_t|\textbf{u}^m_{t-1})$ ;
\item updating step: for each $m=1...M$, compute the incremental function $\alpha_t^m$ from Eq.~\eqref{e:alpha};
 update the weights $w_t^m =  \alpha_{t}^m w_{t-1}^m$,
and  renormalize them as $W_t^m = \frac1{\sum_{m'=1}^M w_t^{m'}} w_t^m$ for each $m=1...M$.  
\end{enumerate}
\end{algorithm}
}}

In the standard PF method, one simply takes $q_0(\-u_0) = \pi(\-u_0)$ and \[q_{t}(\-u_{t}|\-u_{t-1}) = \pi(\-u_{t}|\-u_{t-1}),
 \] 
and as a result, $w_0 =  \pi(\-y_0|\-u_0)$, and the incremental weight function becomes
\begin{equation}
\alpha_{t} (\-u_{0:t}) = \pi(\-y_{t}|\-u_{t}). 
\end{equation}
In the PF algorithm, the variance of the importance  weight $w_t(\-u_{0:t})$ will increase over time, and thus {\color{black}as} the time $t$ increases, the IS weights will become negligibly small for all but one sample, 
an issue known as particle degeneracy. 
To address the issue, a resampling step is often performed to obtain a set of equally weighted particles, the procedure of which 
can be found in, for example,  \cite{SMC,doucet2009tutorial}.
\subsection{The ensemble Kalman filter}
We now discuss the EnKF method which computes a Gaussian approximation of the posterior distribution $\pi(\-u_t|\-y_{0:t})$. 
For simplicity we assume that the observation operator $H_t$ is linear and the observation noise $\bm{\eta}_t$ is Gaussian (these assumptions can be relaxed). 
 Now suppose that
 at time $t$,  the observation noise is $\bm\eta_{t}\sim N(0, R_{t})$ and the prior {\color{black}$\pi(\-u_{t}|\-y_{0:t-1})$} can be approximated by a Gaussian distribution with 
 mean $\tilde{\bm\mu}_{t}$ 
and covariance $\tilde{\Sigma}_{t}$. 
It follows that the posterior distribution {\color{black}$\pi(\-u_{t}|\-y_{0:t})$ }is also  Gaussian and its mean and covariance can be obtained analytically: 
\begin{equation}
{\bm\mu}_t = \tilde{\bm\mu}_t +K_t(\-y_t-H_t\tilde{\bm\mu}_t), \quad {\Sigma}_t = (I-K_tH_t)\tilde{\Sigma}_t , \label{e:postparams}
\end{equation}
where $I$ is the identity matrix and
\begin{equation}
K_t =\tilde{\Sigma}_t H_{t}^T(H_{t}\tilde{\Sigma}_t H_t^T+R_{t})^{-1}
\end{equation}
 is the so-called Kalman gain matrix. 
Moreover, when the prior $\pi(\-u_{t}|\-y_{0:t-1})$ is exactly Gaussian, this formulation becomes the standard Kalman filter. 

In the EnKF method, one avoids computing the mean and the covariance directly in each step. 
Instead, both the prior and the posterior distributions are represented with a set of samples, known as an ensemble. 
Specifically, let $\{\tilde{\-u}_t^m\}_{m=1}^M$ be a set of samples drawn from the prior distribution $\pi(\-u_{t}|\-y_{0:t-1})$,
and we shall compute a Gaussian approximation of $\pi(\-u_{t}|\-y_{0:t-1})$ from the samples. 
Namely we estimate the mean and the covariance of $\pi(\-u_{t}|\-y_{0:t-1})$ from the samples:
\begin{equation}
\tilde{\bm\mu}_t = \frac1M\sum_{m=1}^M {\color{black}\tilde{\-u}_{t}^m}, \quad \tilde{\Sigma}_t={\color{black} \frac1{M-1}\sum_{m=1}^M(\tilde{\-u}_t^m-\tilde{\bm\mu}_t )(\tilde{\-u}_t^m-\tilde{\bm\mu}_t )^T},  \label{e:priorparams}
\end{equation}
and as is mentioned earlier, the prior distribution $\pi(\-u_{t}|\-y_{0:t-1})$ can be approximated by $N(\tilde{\mu}_t, \tilde{\Sigma}_t)$.
 It is not hard to see that  
the posterior distribution is also Gaussian with mean ${\bm\mu}_t$ and covariance ${\Sigma}_t$ given by Eq.~\eqref{e:postparams}.
Moreover it can be verified that the samples 
\begin{equation}
{\-u}_{t}^m =\tilde{\-u}_t^m +K_t(\-y_t-(H_t\tilde{\-u}_{t}^m+\bm\eta^m_t)), \quad \bm\eta_t^m\sim N(0,R_t), 
\end{equation}
 follow the distribution $N({\bm\mu}_t,{\Sigma}_t)$, provided that $\tilde{u}_t^m \sim N(\tilde{\bm\mu}_t, \tilde{\Sigma}_t)$ for all $m=1...M$. 
That is,  $\{\-u_{t-1}^m\}_{m=1}^M$ is the (posterior) ensemble at step $t$.
Given the ensemble $\{\-u_{t-1}^m\}_{m=1}^M$ at time $t-1$, the EnKF algorithm performs the following two steps at time $t$: 
\begin{itemize}
\item prediction step: for each $m=1...M$, draw $\tilde{\-u}_t^m = f_t(\tilde{\-u}_t|\-u^m_{t-1})+\bolde_t^m$ ;
\item updating step:  for each $m=1...M$, compute ${\-u}_{t}^m =\tilde{\-u}_t^m +K_t(\-y_t-H_t\tilde{\-u}_{t}^m{\color{black}-\eta_{t}^m})$.
\end{itemize}
Finally we should note that, as the dynamical model is generally nonlinear, the EnKF method can only provide an approximation of
the true posterior distribution, no matter how large the sample size is, which is major limitation of the method. 

\section{The defensive marginal PF algorithm}\label{Sec:mpf}

\subsection{The marginal particle filter}
As is discussed in Section~\ref{sec:smc}, the standard PF method aims to perform IS for the joint posterior distribution $\pi(\-u_{0:t}|\-y_{0:t})$,
where the dimensionality of the state space grows as $t$ increases. 
On the other hand, in many practical filtering problems, one is often only interested in the marginal posterior distribution at each step, $\pi(\-u_{t}|\-y_{0:t})$, rather than the whole joint distribution. 
This then yields a simple idea: if we perform IS in the marginal space, the dimensionality of the problem is thus fixed and much smaller than that of the joint parameter space. 
For any time $t$, suppose that there is a function defined on the marginal space: $h_t: R^{n_u}\rightarrow R$, and  we are interested in the posterior expectation of $h_t(\-u_t)$:
\[ I = \int h_t(\-u_t) \pi(\-u_t|\-y_{0:t})d\-u_{t} .
\]
We shall construct an IS distribution $q_t(\-u_t|\-y_{0:t})$,
and estimate $I$ as
\begin{equation}
\hat{I}_{IS} =  \frac1M\sum_{m=1}^M h_t(\-u_{t}^m) w_t(\-u_{t}^m), \label{e:margis}
\end{equation}
where $\-u_t^m$ are drawn from $q_t(\-u_t)$ and 
$w_t(\-u_t^m) = \pi(\-u_t|\-y_{0:t})/q_t(\-u_t)$. 
The key issue here is certainly to find a good IS distribution $q_t(\-u_t)$, and ideally this IS distribution should approximate the marginal posterior $\pi(\-u_t|\-y_{0:t})$. 
We first note that a special choice of the IS distribution is 
\begin{equation}
q_t(\-u_t) = \pi(\-u_t|\-y_{0:t-1}), \label{e:q4pf}
\end{equation}
and it should be clear that the associated weight becomes $w_t(\-u_t) = \pi(\-y_t|\-u_t)$ and the algorithm is essentially equivalent to the standard PF method.  
In  \cite{klaas2005toward}, a  kernel-based IS distribution is suggested:
\[ q_t(\-u_t) = \sum_{m=1}^M w_{t-1}(\-u^m_{t-1}) Q_m(\-u_t|\-u^m_{t-1}),\]
where each $Q_m$ is obtained using a weighted Kernel density estimation (KDE) method. 
As a result the method requires to perform a weighted KDE procedure at each time step, which can be computationally intensive even with some fast KDE algorithms (e.g. the dual-tree methods). We discuss an alternative approach to construct the IS distribution in the next section.

\subsection{ The EnKF-based IS distribution}
When the marginal posterior is close to Gaussian, 
the EnKF method can compute a good IS distribution in a very efficient manner. 
Loosely speaking, at a given time, we first compute an ensemble of the marginal posterior distribution using the EnKF scheme, estimate the associated Gaussian approximation from the ensemble, and use it as the IS distribution in the marginal PF. 
Specifically, let $\{\-u_t^m\}_{m=1}^M$ be the posterior ensemble at time $t$ obtained with the EnKF formulation, we use the following procedure to compute the IS distribution:

\begin{algorithm}[H]
	\caption{Estimating the IS distribution from the ensemble} \label{alg:enkf}
	\begin{enumerate}
		\item estimate the mean and covariance from the posterior ensemble $\{{\color{black}\textbf{u}}_t^m\}_{m=1}^M$:
		\begin{equation}
		\mu_\mathrm{En}={\color{black}\frac1M\sum_{m=1}^M{\color{black}\textbf{u}}_t^m},\quad \Sigma_\mathrm{En}={\color{black}\frac{1}{M-1}\sum_{m=1}^M({\color{black}\textbf{u}}_t^m-\mu_\mathrm{En})({\color{black}\textbf{u}}_t^m-\mu_\mathrm{En})^T};
		\end{equation}
		let $q'_{\color{black}\mathrm{EnKF}}({\color{black}\textbf{u}}) = N(\mu_\mathrm{En},\Sigma_\mathrm{En})$;
		\item draw $M$ samples ${\color{black}\textbf{u}}_t^1,...,{\color{black}\textbf{u}}_t^M$ from {\color{black}$q'_\mathrm{EnKF}$}, and compute the weight of each sample, and renormaliz it:
		\[
		w_t^m= \frac{\pi({\color{black}\textbf{u}}_t^m|\-y_{0:t})}{q'_\mathrm{EnKF}({\color{black}\textbf{u}}_t^m)},\ {\color{black} W_t^m = \frac{w_t^m}{\sum_{m'=1}^M w_t^{m'}}} ;
		\]
		\item estimate the mean and covariance of the weighted ensemble $\{({\color{black}\textbf{u}}_t^m,W_t^m)\}$:
		\begin{equation}
		\bm\mu_\mathrm{updated} = {\color{black}\sum_{m=1}^MW_t^m{\color{black}\textbf{u}}_t^m},\quad \Sigma_\mathrm{updated}={\color{black}\sum_{m=1}^MW_t^m({\color{black}\textbf{u}}_t^m-\bm\mu_\mathrm{updated})({\color{black}\textbf{u}}_t^m-\bm\mu_\mathrm{updated})^T};
		\end{equation}
		let ${\color{black}q}_\mathrm{EnKF}=N(\bm\mu_\mathrm{updated},\Sigma_\mathrm{updated})$.
	\end{enumerate}
\end{algorithm}
It is worth noting that the EnKF ensemble does not exactly follow the posterior distribution, 
and thus we choose not to use directly $q'_\mathrm{EnKF}$ in Algorithm~\ref{alg:enkf}, i.e., the Gaussian approximation estimated from the EnKF ensemble, 
as the IS distribution.
Instead, we introduce additional steps (Steps 2 and 3 in Algorithm~\ref{alg:enkf}), in which we first generate a weighted ensemble according to the true posterior, and then
update the Gaussian approximation according to this weighted ensemble.
By doing so we ensure that the Gaussian approximation is constructed with respect to the true posterior ensemble.  

\subsection{The defensive marginal particle filter}
As has been discussed earlier, when strong nonlinearity is present, the true posterior can no longer be approximated by a Gaussian distribution.
In this case the IS distribution obtained with the EnKF method may deviate significantly from the true marginal posterior,
 leading to erroneous estimates of the states.  
To address the issue, we use the idea of \emph{multiple importance sampling} (MIS), also known as the deterministic mixture~\cite{owen2000safe}. 
That is, to prevent the failure of the IS distribution computed with the EnKF method, one uses a mixture of  the Gaussian approximation computed by EnKF
and a safe distribution, which in our case is the standard PF distribution, yielding, 
\begin{equation}
\label{equ:DIS}
q_t(\-u_t|a) = aq_\mathrm{EnKF}(\-u_t) + (1-a)q_\mathrm{PF}(\-u_t), 
\end{equation}
where $q_\mathrm{EnKF}$ is the Gaussian distribution computed with the EnKF procedure described above, $q_\mathrm{PF} $ is the distribution given by Eq.~\eqref{e:q4pf}, which, as discussed earlier, is equivalent to the standard PF,
and $a\in[0,1]$ is the weight of the EnKF component, which will be referred 
to as the weight parameter hereafter. 
This mixture distribution~\ref{equ:DIS} is the defensive IS scheme proposed in \cite{hesterberg1995weighted}.
If this mixture IS distribution is used in Eq.~\eqref{e:margis} to estimate $I$, 
 the estimation variance of the mixture is bounded by \cite{hesterberg1995weighted}:
\[\sigma^2_\mathrm{DIS} \leq \frac1{1-a}\sigma^2_\mathrm{PF} + \frac{1-a}{a} I^2,\]
with $\sigma^2_\mathrm{PF}$ being the estimation variance of $q_\mathrm{PF}$, 
regardless of how large the estimation variance of $q_\mathrm{EnKF}$ is. 
The MIS scheme is very similar to the mixture IS distribution in Eq.~\eqref{equ:DIS}, except that it generates
a {fixed number} of samples from each component: namely $aM$ samples from $q_\mathrm{EnKF}$ and $(1-a)M$ from $q_\mathrm{PF}$,
and hence the name deterministic mixture. It has been shown that the use of MIS often 
yields more accurate and robust estimates than the standard mixture IS distribution~\cite{hesterberg1995weighted,owen2000safe},
and so here we adopt the MIS method as our defensive scheme. 

An important issue here is how to compute the IS weight of each sample in the MIS scheme.
In \cite{veach1995optimally}, the authors recommend the so-called balance heuristic weight: 
\begin{equation}
\label{equ:DW}
w_t(\-u_t) = \frac{\pi(\-u_{t}|\-y_{0:t})}{aq_\mathrm{EnKF} + (1-a)q_\mathrm{PF}}=\frac1{\frac{a}{w_\mathrm{EnKF}} + \frac{(1-a)}{w_\mathrm{PF}}},
\end{equation}
where 
\begin{equation}
w_\mathrm{EnKF} = \frac{ \pi(\-y_t|\-u_t)\pi(\-u_t|\-y_{0:t-1})}{\pi({\color{black}\-y_t|\-y_{0:t-1}})q_\mathrm{EnKF}(\-u_{t})}, \quad w_\mathrm{PF}= \pi(\-y_t|\-u_t){\color{black}/\pi(\-y_t|\-y_{0:t-1})}.\label{e:wts}\end{equation}
Computing $w_\mathrm{PF}$ is rather straightforward, but computing $w_\mathrm{EnKF}$ involves the evaluation of the integral:
\begin{equation}
\pi(\-u_t|\-y_{0:t-1}) = \int \pi(\-u_t|\-u_{t-1})\pi(\-u_{t-1}|\-y_{0:t-1}) d\-u_{0:t-1}.
\end{equation}
In practice, this integral is approximated by 
\begin{equation}
\label{e:wcomp}
\pi(\-u_t|\-y_{0:t-1}) \approx \sum_{m=1}^M w_{t-1}^m\pi(\-u_t|\-u_{t-1}^m), 
\end{equation}
where  $\{\-u_{t-1}^m\}_{m=1}^M$ are the samples generated in the previous step and $w_{t-1}^m$ is the associated weight of each sample $\-u^m_{t-1}$~(namely, the weighted ensemble
$\{(\-u_{t-1}^m,w_{t-1}^m)\}_{m=1}^M$ follows the distribution $\pi(\-u_{t-1}|\-y_{0:t-1})$).
We note that, as evaluating Eq.~\eqref{e:wcomp} {\color{black}requires summing} over $M$ particles,
computing all the weights is of $M^2$ complexity, which can be highly intensive when the number of particles is large. 
However, by using the fast multipole method~\cite{greengard1987fast} one can reduce the computational cost to 
$M\log M$ (see \cite{klaas2005toward} for more discussions).
Another important matter in the proposed method is to determine the value of the weight parameter $a$,
which is discussed in Section~\ref{sec:weight}. 
We hereby provide the complete defensive marginal PF (DMPF) algorithm in Algorithm~\ref{alg:DMPF}.
\IncMargin{1em} 
\begin{algorithm}   
	\caption{The DMPF algorithm\label{alg:DMPF}}	
	\SetAlgoNoLine 
	
	
	At $t=0$:\\
	{Prediction: sample $\{\tilde{{\color{black}\textbf{u}}}_0^m\}_{m=1}^M$ from $\pi_0\-(\cdot)$;\\
		Updating: ${\-u}_{0}^m =\tilde{\-u}_0^m +K_0(\-y_0-H_0\tilde{{\color{black}\textbf{u}}}_{0}^m{\color{black}-\eta_{t}^m})$ for $m=1...M$;\\
		Compute $q_\mathrm{EnKF}$ using Algorithm~\ref{alg:enkf} and particles~$\{{\color{black}\hat{\textbf{u}}}_0^m\}_{m=1}^M$;\\
		Draw $M$ particles from $q_0$ from Eq.~\eqref{equ:DIS} for $t=0$, and compute the weights using Eq.~\eqref{e:wts}, yielding
		$\{({\color{black}\textbf{u}}_0^m, w^m)\}_{m=1}^M$;\\}
	\For{t=1...T}
	{Prediction: for each $m=1...M$, draw $\tilde{{\color{black}\textbf{u}}}_t^m = f_t(\tilde{{\color{black}\textbf{u}}}_t|\-u^m_{t-1})+\bolde_t^m$;\\
		Updating: ${{\color{black}\textbf{u}}}_{t}^m =\tilde{{\color{black}\textbf{u}}}_t^m +K_t(\-y_t-H_t\tilde{{\color{black}\textbf{u}}}_{t}^m)$ for $m=1...M$;\\
		Compute $q_\mathrm{EnKF}$ using Algorithm~\ref{alg:enkf} and particles~$\{{\color{black}\hat{\textbf{u}}}_t^m\}_{m=1}^M$;\\
		Estimate the weight parameter $a$ by solving the optimization problem~\eqref{e:minvar};\\
		Draw $M$ particles from $q_t$ given by Eq.~\eqref{equ:DIS}, and compute the weights using Eq.~\eqref{e:wts}, yielding
		$\{({\color{black}\textbf{u}}_t^m, w^m)\}_{m=1}^M$;\\}
	
\end{algorithm}
\DecMargin{1em}

\subsection{Optimizing the weight parameter}\label{sec:weight}
It is highly important to choose an appropriate value for the weight parameter $a$ in the DMPF algorithm. 
In previous works on MIS, a fixed $a$ is often used, and in particular, it is suggested in \cite{hesterberg1995weighted} that the parameter should be chosen between $0.1-0.5$.
In this section, we provide a method that can automatically determine the value of $a$ at each time step. 
The method can assign more weight to the EnKF component if the posterior is close to Gaussian
and to the PF component otherwise. 

As is well known in the PF literature, the optimal IS distribution should yield equal IS weights, i.e., $w_t(\-u_t)=1$ for all $\-u_t$,
which is usually not possible in practice. 
Nevertheless, this gives us the idea that the variance of the weight function associated with distribution $q_t(\-u_t|a)$ should be as small as possible, and so we can determine the value of $a$ by minimizing the variance of $w_t(\-u_t,a)$ (here we use the notation $w_t(\-u_t,a)$ to emphasize the dependence of $w_t$ on $a$):
\begin{eqnarray}
\min_{a\in[0,1]} \mathrm{Var}_{q_t(\-u_t|a)}[w_t(\-u_{t},a)] &=& \E_{q_t(\-u_t|a)}[(w_t(\-u_t,a)-1)^2]\\
&=& \int (w_t(\-u_t,a)-1)^2 q_t(\-u_t|a) d\-u_t.
 \end{eqnarray}
Optimizing the variance directly is usually not feasible, and so a natural idea is to optimize its sample-average approximation:
\[ \min_{a\in[0,1]}\frac1M\sum_{m=1}^M(w_t(\-u_t^m,a)-1)^2,
\]
where $\{\-u_t^m\}_{m=1}^M$ are drawn from distribution $q_t(\-u_t|a)$. 
However, it is actually undesirable to use this sample average approximation,  in that, whenever $a$ is updated, we will have to
generate new samples and compute the associated weights, 
which, as is discussed earlier, is of $M^2$ complexity.
To address the issue,  we choose an default value of $a$, say $a_0$ (in this work we choose $a_0=0.5$), and apply an IS simulation with distribution $q_t(\-u_t|a_0)$ to estimate the variance: namely,
\begin{equation}
 \min_{a\in[0,1]}\frac1M\sum_{m=1}^M(w_t(\-u_t^m,a)-1)^2 w_t({\color{black}\-u_t^m},a_0),\label{e:minvar}
\end{equation}
where samples are drawn from $q_t(\-u_t|a_0)$.
Solving the optimization program~\eqref{e:minvar} does not affect the computational efficiency of the algorithm much and the reason is two-fold.
First this optimization problem is rather easy to solve as it is only a one-dimensional problem, 
specially as our method does not require high accuracy in the solution here.
Secondly by design, solving this optimization problem does not require additional evaluations of the dynamical model $f_t$ which is usually the most computationally intensive part in the problem, or the weights that are of $M^2$ computational complexity. 

\section{Numerical examples}\label{Sec:examples}
In this section we provide three numerical examples to compare the performance of the proposed DMPF method, with EnKF, PF,
as well as two existing methods that combine EnKF and PF: {\color{black}the weighted EnKF (WEnKF) method in \cite{papadakis2010data} and the Ensemble Kalman Particle Filter (EnKPF) in \cite{frei2013bridging}. 
Here we provide a brief introduction to the two methods. The WEnKF method uses the EnKF to construct
a proposal distribution in the PF framework, yielding weighted samples~\cite{papadakis2010data}.
The EnKPF uses the “progressive correction” idea and introduces an ``intermediate'' posterior distribution at each time step; it then updates from the prior to the intermediate distribution using EnKF and from the intermediate 
distribution to the complete posterior using PF~\cite{frei2013bridging}.
We note here that the EnKPF method is particularly similar to that propose in the present work, 
as it also provides a continuous transition indexed by $a\in [0,1]$ ($\gamma$ in \cite{frei2013bridging}) between EnKF and PF, which is determined 
automatically.}
\subsection{Bernoulli model}\label{sec:bern}
Our first example is the Bernoulli equation,   
	\begin{equation}
	\frac{dx}{dt} -x=-x^3,\ \ \ x(t_0)=x_0, 
	\end{equation}
	which admits an analytical solution, 
	\begin{equation}
	x(t)={M}(x_0,\Delta t)=x_0\times (x_0^2+(1-x_0^2)e^{-2\Delta t})^{-1/2},
	\end{equation}
	where $\Delta t = t-t_0$. 
	Here for simplicity we use the analytical solution to construct the discrete-time model: 
	\begin{equation} 
		\label{eq:Ber}
		\begin{array}{ll}
		x_0 \sim N(\mu_0,\sigma_0),\\ 
		x_k ={M}(x_{k-1},\Delta t)+\xi_k,\\
		y_k=x_k+\eta_k,
		\end{array}
		\end{equation}
where $\xi_k$ and $\eta_k$ are the model and observation noise respectively. 
In this example we set $x_0\sim N(-0.1, 0.2^2)$, {\color{black}$\Delta t=0.3$ }and the total number of steps to be $40$. 
Moreover, we assume that both $\xi_k$ and $\eta_k$ follow zero-mean Gaussian distributions with standard deviation {\color{black}$0.01$ and $0.8$}. 
This is an often used example with strongly non-Gaussian posteriors~\cite{apte2007sampling,stordal2011bridging}. 

We generate a true state and the associated data points from the model~\eqref{eq:Ber}, which are shown in Fig.~\ref{f:data_bern}.
We first draw $5\times10^5$ particles using the standard PF method to represent the true posteriors. {\color{black} 
We then perform the five aforementioned methods to estimate the states, each with $10^4$ particles.
In EnKPF, the constrained diversity $\tau$ used to determine the weights (see \cite{frei2013bridging} for details)  is taken to be $[0.9,1]$.}
We compare the posterior means and variances computed by all the methods in Fig.~\ref{f:bern}.
One can see from the plots that,  the DMPF method yields results in a good agreement with the truth, while those of the EnKF significantly depart as the time proceeds. 
The  results of the EnKPF method are better than EnKF, but still deviate evidently from those of PF and DMPF. 
The poor performance of the EnKF method in this example can be understood by examining the posterior distributions:
in Fig.~\ref{f:pdf_bern}, we plot the posterior distributions computed by all methods at steps $5$ and $10$ respectively, where the distributions of the PF and DMPF methods are obtained by performing a kernel density estimation with the particles.
 As one can see here, while at $k=5$ the EnKF approximation remains rather close to the true posterior distribution, 
it significantly deviates from the the true posterior at $k=10$ because of the cumulation of the non-Gaussianity as time increases. 
We also implement the WEnKF, the results of which are highly unstable and so are not plotted in these figures.

{\color{black}Moreover, recall that in both DMPF and EnKPF  a scalar parameter $a\in[0,1]$ is used to describe 
the relative strength of the EnKF and the PF components in the algorithm, and in both algorithms, the parameter is automatically determined. 
We plot the parameter $a$ estimated in both methods at each time step in Fig.~\ref{f:bern_alpha}. 
It can be seen from the figure that, in most of the steps the weight parameter $a$ is close to $0$ in both methods, suggesting that both methods detect that the EnKF approximation is not a good approximation to the posterior in most steps and and are
able to choose  suitable values for $a$ accordingly. }
It should also be noted here that the EnKF method usually employs a rather small number of particles, e.g. several hundreds,
and here we intentionally uses a rather large number of particles to demonstrate that the large bias in the EnKF approximation 
can not be reduced by increasing the number of samples. 
In summary, this example demonstrates that, in presence of strong  non-Gaussianity, the Gaussian approximation computed by 
the EnKF method may fail completely, 
while our DMPF method can nevertheless produce accurate posterior estimates. 

\begin{figure}
\centerline{\includegraphics[width=.55\textwidth]{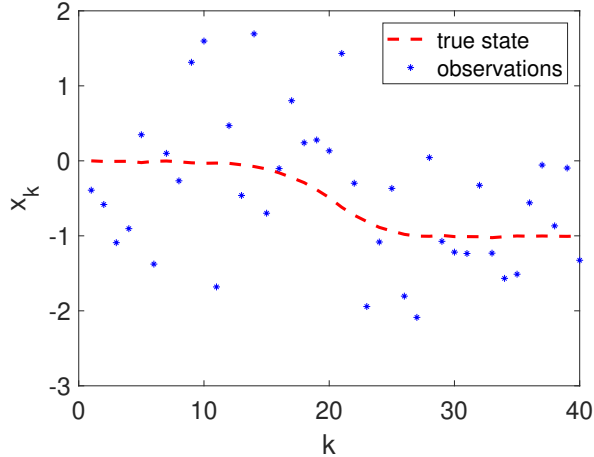}} 
\caption{The true state (dashed lines) and the simulated observations (dots) of the Bernoulli model.}\label{f:data_bern}
\end{figure}	

\begin{figure}
\centerline{\includegraphics[width=.55\textwidth]{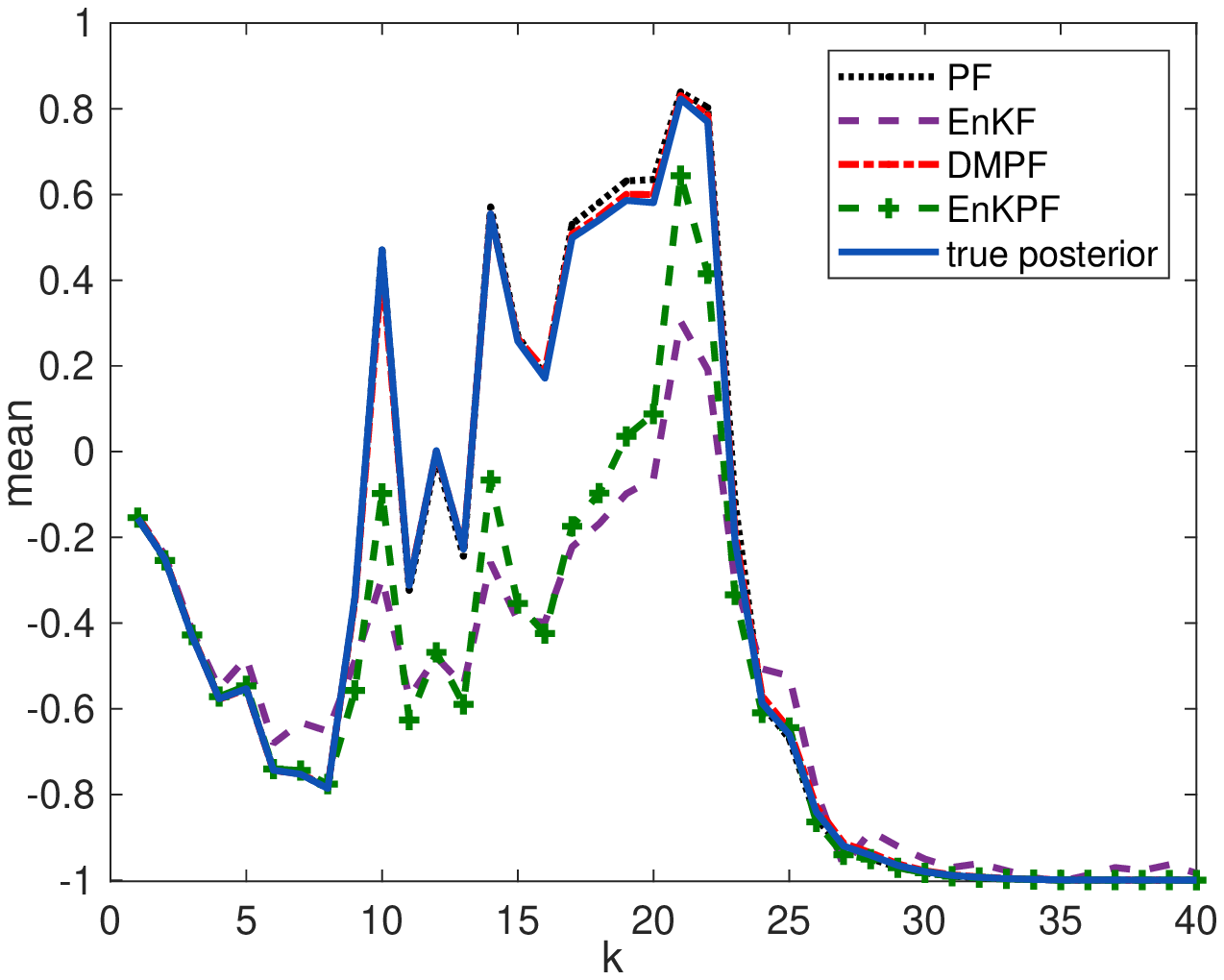}\includegraphics[width=.55\textwidth]{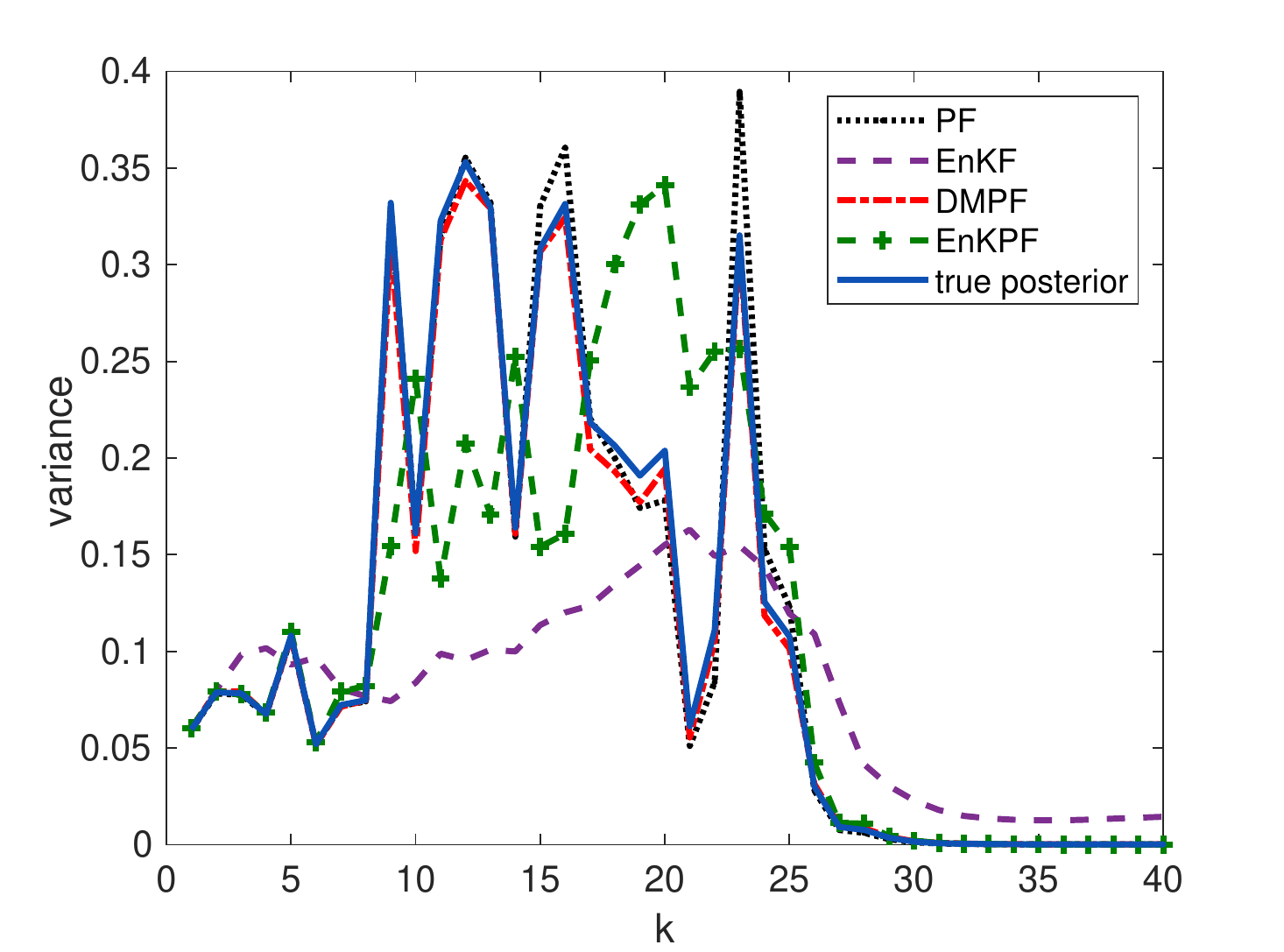}} 
\caption{Left: the posterior means computed by the different methods. Right: the posterior variances computed by the different methods.}\label{f:bern}
\end{figure}	


\begin{figure}
\centerline{\includegraphics[width=.55\textwidth]{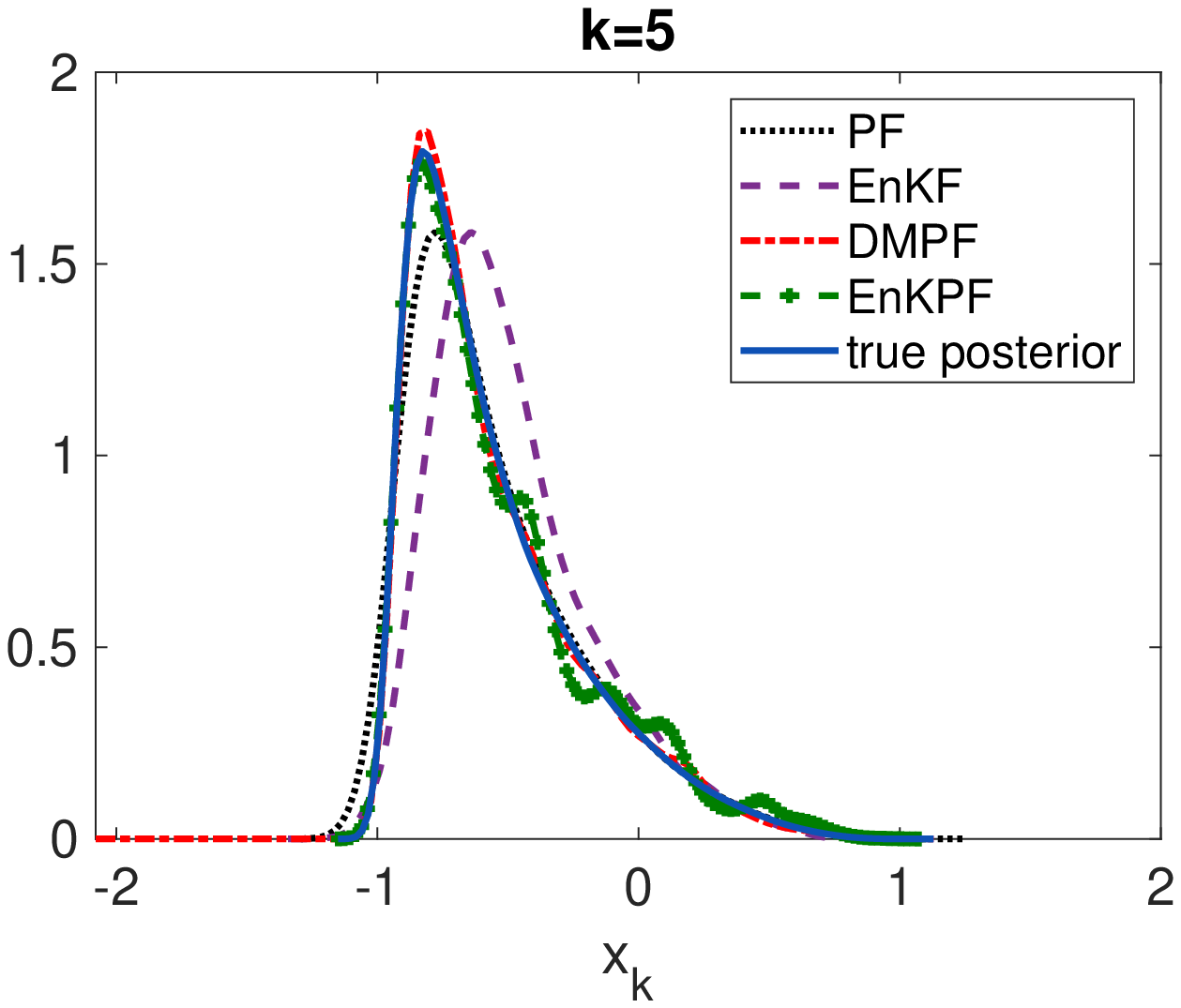}\includegraphics[width=.55\textwidth]{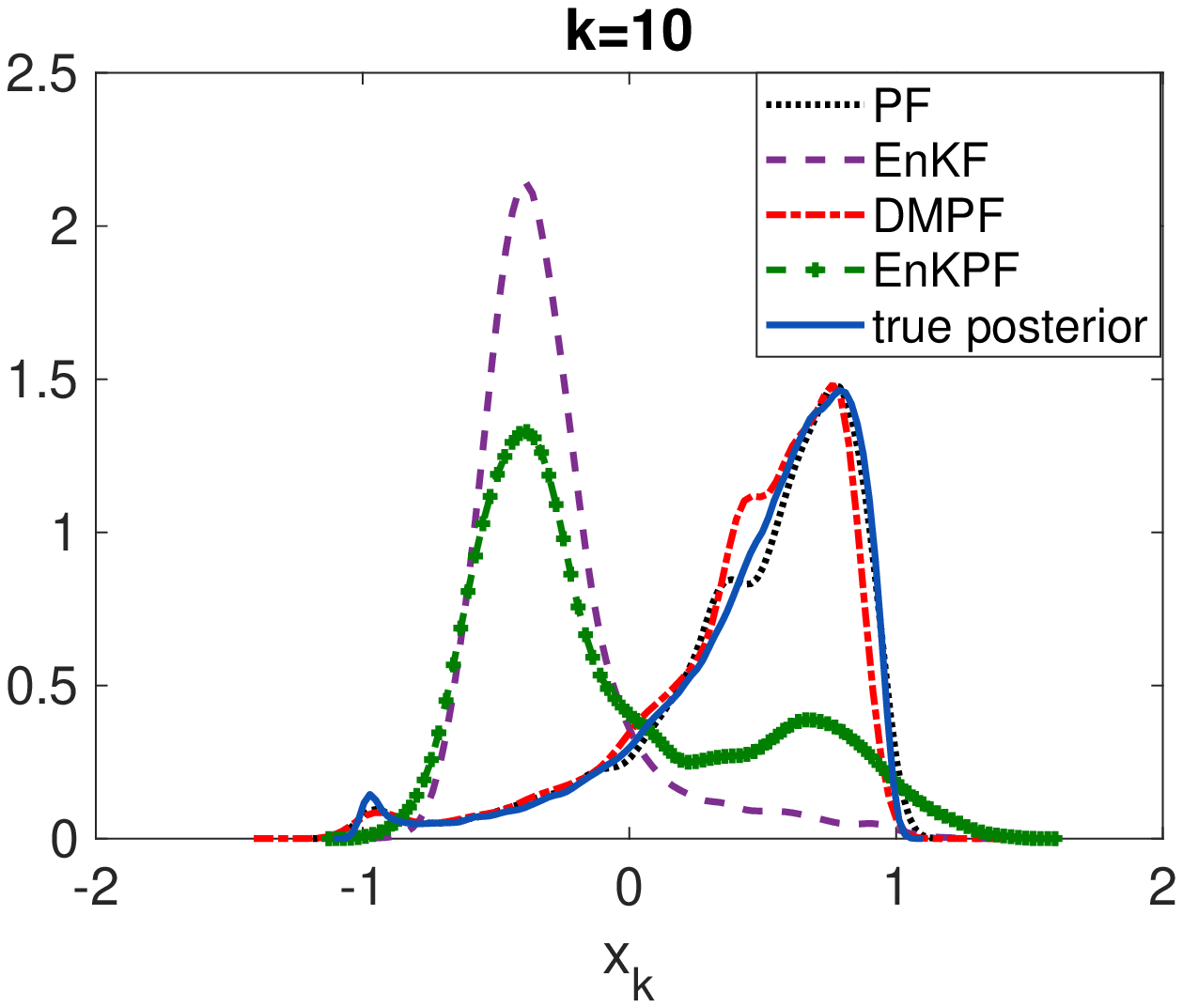}} 
\caption{Left: the posterior distributions at $k=5$. Right: the posterior distributions at $k=10$.
In both plots, the solid lines are the true posteriors and the dashed ones are the EnKF approximations.}\label{f:pdf_bern}
\end{figure}

\begin{figure}
\centerline{\includegraphics[width=.55\textwidth]{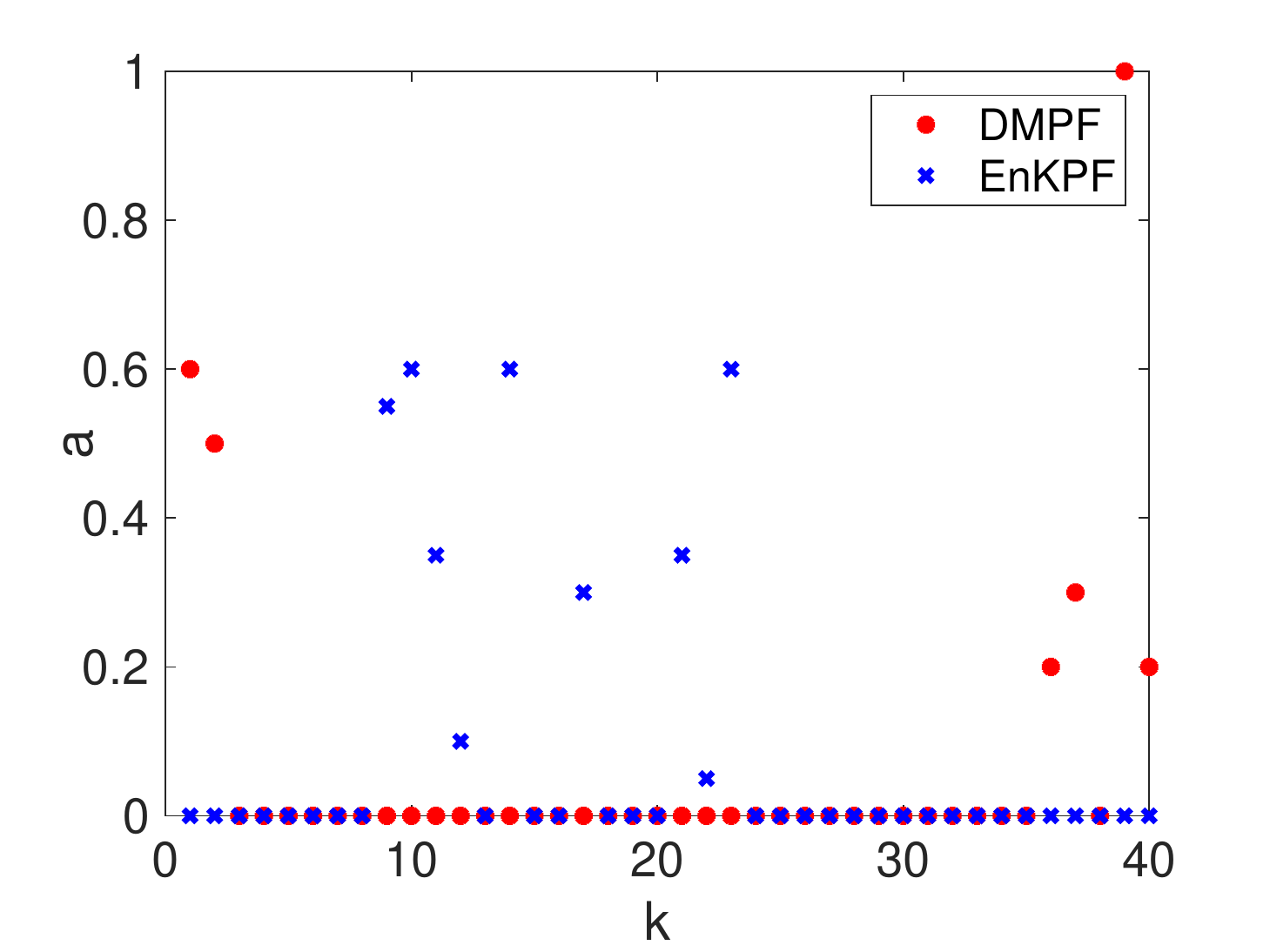}} 
\caption{The weight parameter $a$ computed  at each time step in DMPF and EnKPF in 
the Bernoulli model.}\label{f:bern_alpha}
\end{figure}	

\subsection{Lorenz 63 system}
Our second example is the classical Lorenz 63 system~\cite{lorenz1963deterministic}, an often used benchmark problem for testing data assimilation algorithms. 
Specifically the system is described by, 
\begin{equation} \label{eq:lorenz}
	\left\{\begin{array}{ll}
	\dot{x} &=\sigma(y-x),\\
	\dot{y} &=x(\rho-z)-y,\\
	\dot{z} &=xy-\beta z.
	\end{array}
	\right.
	\end{equation}
For simplicity, we consider a discrete-time version of the system with additive noise: 
\begin{equation} \label{eq:discretelorenz}
	\left\{\begin{array}{ll}
	x_{t+1}&=x_t+\sigma(y_t-x_t)\Delta t +\xi_t^x,\\
	y_{t+1}&=y_t+(x_t(\rho-z_t)-y_t)\Delta t+\xi_t^y,   \\
	x_{t+1}&=z_t+(x_ty_t-\beta z_t)\Delta t+\xi_t^z,
	\end{array}
	\right.
	\end{equation}
	where $\Delta t$ is the discrete-time step size. 
Here we assume that the model noise $\xi_t^x,\,\xi_t^y$ and $\xi_t^z$ are all i.i.d zero-mean Gaussian with  standard deviation $\sigma_\xi$.
Moreover at each time $t$ the observed data is taken to be $x^d_t = x_t +\eta_t^x$, $y^d_t = y_t +\eta_t^y$ and $z^d_t = z_t +\eta_t^z$,
where the observation noise $\eta_t^x$, $\eta_t^y$ and $\eta_t^z$ are once again assumed to be i.i.d. zero-mean Gaussian with standard deviation $\sigma_\eta$.

In our numerical tests we take the parameters to be $\sigma=10$, $\rho=28$, $\beta=8/3$, {\color{black}$\Delta t= 0.03$ and $150$ steps}, and the initial condition to be 
\[[x_0,\,y_0,\,z_0]=[1.51,\, -1.53,\, 25.46].\]
The noise standard deviations are $\sigma_\epsilon={0.5}$ and $\sigma_\eta ={ 1}$. 
In this example we also use a simulated true state and generate noisy data from it, where both of them are shown in Fig.~\ref{f:lorenz_data}. 
We use this example to quantitatively compare the performance of these methods.
We repeat the simulations 1000 times for all methods with $10^4$ particles each time, 
and so we can examine the statistical behavior of the  methods.  
Also, as a comparison reference, we also perform a PF with $5\times10^5$ particles to represent the true posteriors. 
In Fig.~\ref{f:lorenz_alpha}, we show the weight parameter $a$ computed in one of the DMPF and EnKPF simulations, and one can see from the figure that, 
unlike the first example,  the parameter $a$ calculated in DMPF is close to $1$ in most of steps, indicating that 
the posteriors in those steps can be well approximated by the EnKF approximation.
  
To quantitatively compare the performance of the methods, we compute the root mean squared error (RMSE) for posterior mean and variance: namely, 
in the $j$-th simulation, we estimate the posterior mean $(\hat{\-x}_{j},\hat{\-y}_{j},\hat{\-z}_{j})$, and the posterior variance $(\-V^x_{j},\-V^y_{j},\-V^z_j)$,
and the RMSE is then calculated as, 
\begin{eqnarray*}
\mathrm{RMSE}_\mathrm{mean}&=&\frac1{1000}\sum_{j=1}^{1000} \left[(\hat{\-x}_j-\hat{\-x})^2+(\hat{\-y}_j-\hat{\-y})^2+(\hat{\-z}_j-\hat{\-z})^2\right]^{\frac12},\\
\mathrm{RMSE}_\mathrm{var}&=&\frac1{1000}\sum_{j=1}^{1000} \left[(\-V^{x}_j-\-{\color{black}\-V^{x}})^2+(\-V^y_j-\-V^y)^2+(\-V^z_j-\-V^z)^2\right]^{\frac12},
\end{eqnarray*}
where  $(\hat{\-x},\hat{\-y},\hat{\-z})$, and  $(\-V^x,\-V^y,\-V^z)$ are the true values of the posterior mean and variance~(computed with $5\times10^5$ PF particles). 
It should be clear that the RMSE is a measure of the estimation error: smaller RMSE usually indicates lower estimation error. 
In Fig.~\ref{f:lorenz_rmse}, we plot the RMSE of the posterior mean (left) and variance (left) for all methods,
and in Table~\ref{tab:lorenz_table} we show the RMSE averaged over all time steps. 
{\color{black}Some conclusions can be drawn from the figures and the table.
First the EnKF performs significantly better than PF, and this can be understood  
as that the posteriors in this problem can be well approximated by the EnKF (Gaussian) approximation, and the PF method does not 
take advantage of that.
On the other hand, the DMPF and the EnKPF methods perform similar to the EnKF, suggesting that
both methods can detect the fact that EnKF approximates the posteriors well and adjust 
the algorithm accordingly
to take advantage of it. 
}

\begin{figure}
\centerline{\includegraphics[width=.85\textwidth]{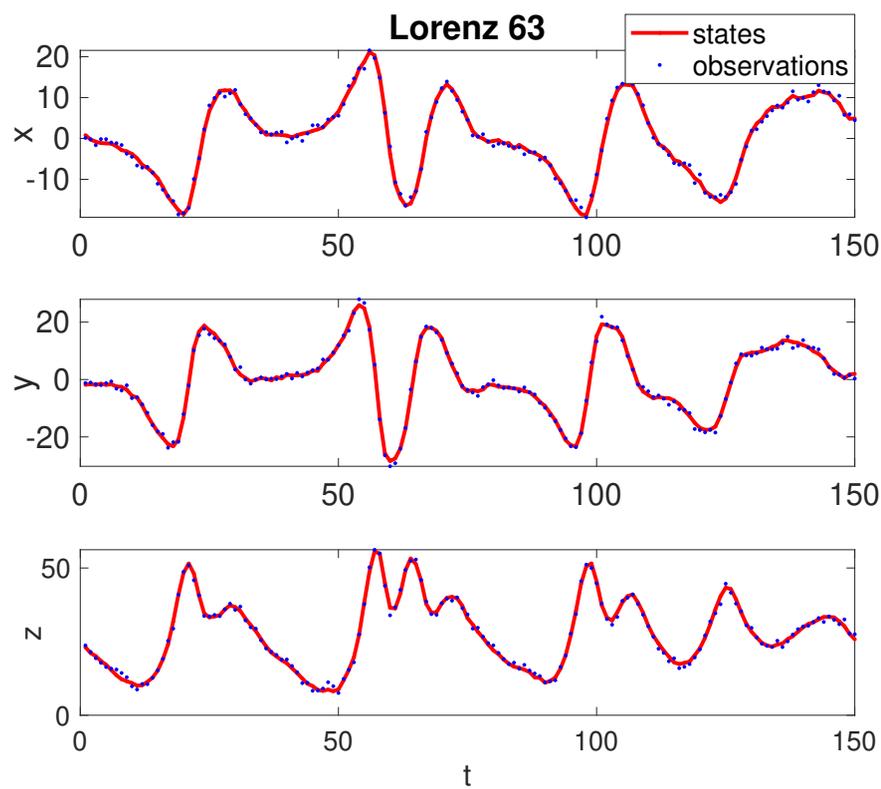}} 
\caption{The true state (dashed lines) and the simulated observations (dots) of the Lorenz 63 model.}\label{f:lorenz_data}
\end{figure}

\begin{figure}
\centerline{\includegraphics[width=.55\textwidth]{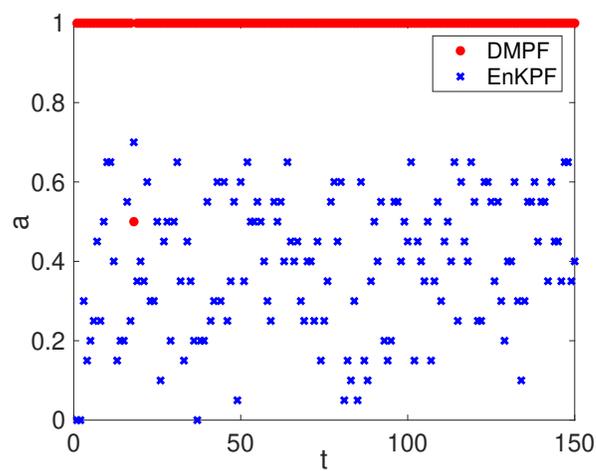}} 
\caption{{\color{black} Parameter $a$ in the DMPF and the EnKPF methods for the Lorenz 63 model.}}\label{f:lorenz_alpha}
\end{figure}

\begin{figure}
\centerline{\includegraphics[width=.55\textwidth]{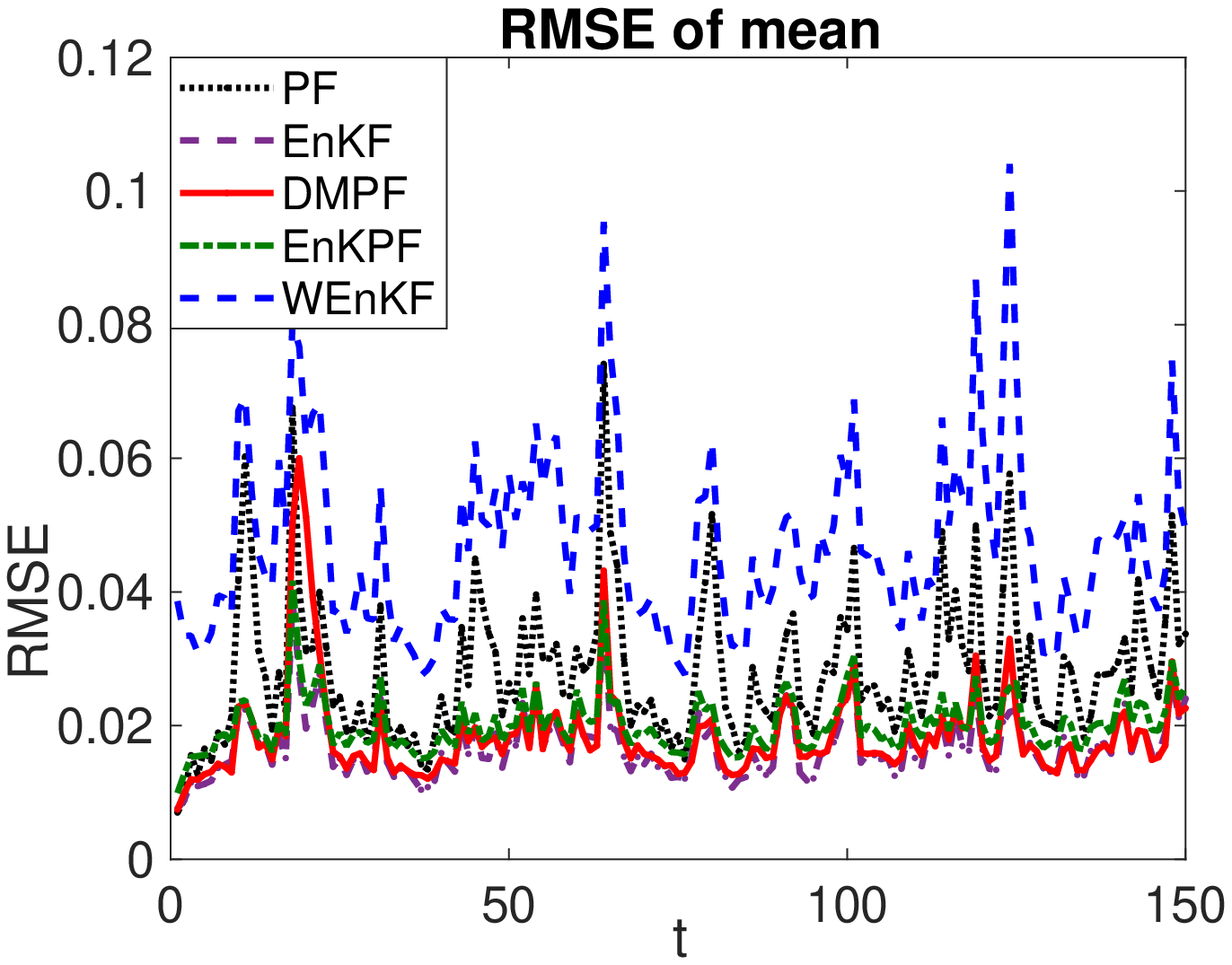}
\includegraphics[width=.55\textwidth]{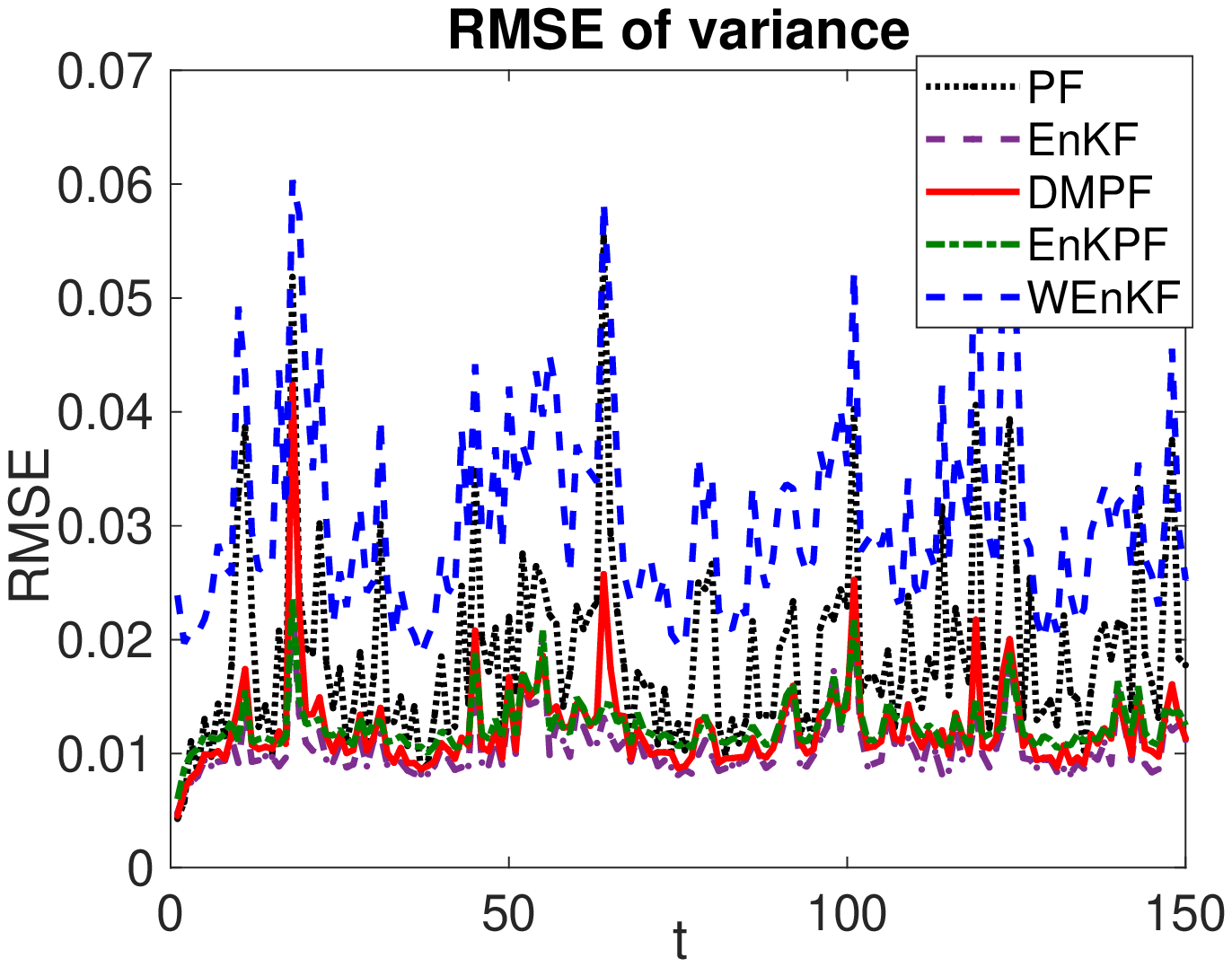}} 
\caption{The RMSE of the posterior mean (left) and of the posterior variance (right) for the Lorenz 63 model.}\label{f:lorenz_rmse}
\end{figure}

\begin{table}[htbp]
	\caption{RMSE averaged over all time steps in the Lorenz 63 model.}
	\label{tab:lorenz_table}
	\centering
	\begin{tabular}{|c|c|c|} \hline
		- & \bf mean & \bf variance \\ \hline
		\bf PF  & 0.028 & 0.019 \\ 
		\bf EnKF  & 0.017 & 0.010\\ 
		\bf DMPF & 0.018 & 0.012 \\ 
		\bf EnKPF & 0.020 & 0.012\\  
		\bf WEnKF & 0.047 & 0.031 \\ \hline
	\end{tabular}	
\end{table}


\subsection{Localization of a car-like robot}
Finally we consider a real-world problem, in which the position of a remotely controlled car-like robot is inferred from the on-board GPS data. 
The kinematic model of the car-like robot is described by the following non-linear system~\cite{klaas2005toward}:
\begin{equation}
\label{eq:carlike}
\begin{array}{ll}
\dot{x} =& v\cos(\theta),\\
\dot{y} =& v\sin(\theta),\\
\dot{\theta} =& \frac{v}{L}\tan(\phi),\\
\dot{\phi} =& \omega,
\end{array}
\end{equation}
where $(x,y)$ are the position coordinates of the vehicle, $L$ is its length,   $\theta$ is the steering orientation angle, {\color{black}$\phi$} is the front wheel orientation angle,
and $v$ and $\omega$ are the linear and angular velocities respectively. 
The schematic illustration of the model is shown in Fig.~\ref{f:robot}.
 In this problem, we assume the linear and the angular velocities  $v$ and $\omega$ are controlled as follows: 
{\color{black}
\begin{equation}
\label{eq:control}
\begin{array}{ll}
v =0.7|\sin(t)|+0.1,\\
\omega =0.08\cos(t).
\end{array}
\end{equation}
The discrete-time version of the model  is described by:
\begin{equation}
\label{eq:discritcarlike}
\left(\begin{array}{ll}
x_{t+1}\\y_{t+1}\\\theta_{t+1}\\\phi_{t+1}
\end{array}\right)=\textit{M}(t,\left(\begin{array}{ll}
x_{t}\\y_{t}\\\theta_{t}\\\phi_{t}
\end{array}\right))+\left(\begin{array}{ll}
\epsilon_x\\\epsilon_y\\\epsilon_{\theta}\\\epsilon_{\phi}
\end{array}\right)
\end{equation}
where $\textit{M}$  represents the standard fourth-order Runge-Kutta solution of Eq.~\eqref{eq:carlike} with $\Delta t=0.05$.
In Eq.~\eqref{eq:discritcarlike}, $\epsilon_x$, $\epsilon_y$, $\epsilon_{\theta}$ and $\epsilon_{\phi}$ are the errors in the state process.
In particular all these errors are taken to be zero mean Gaussian 
with standard deviation $0.3$.}
On the other hand, the GPS makes measurements of the pose $(x,\,y,\,\theta)$ of the vehicle, and specifically 
these measurements are taken to be
\[
\hat{x} = x+\eta_x,\quad \hat{y}=y+\eta_y,\quad \hat{\theta} =\theta+\eta_\theta,\]
where $\eta_x$, $\eta_y$ and $\eta_\theta$ are the observation noise following $N(0,0.3^2)$.  
We shall estimate $x$, $y$, $\theta$ and $\omega$ from these measurements for {\color{black}a time period $T=5$ that is discretized into $100$ steps.} 
The true states of the system are randomly generated and the measurement data are simulated from the generated true states using the prescribed model;
both the true states and the associated measurements are plotted in Fig.~\ref{f:data_car}. 
We emphasize here that no observations are made on the front-wheel angle $\phi$ and so only the true states of it are plotted in Fig.~\ref{f:data_car}.  The constrained diversity $\tau$ in EnKPF is taken to be $[0.25, 0.5]$.

We implement all the five methods to estimate the states of the four parameters in this problem, each with $10^4$ particles.
Once again we repeat the simulations of each method for 1000 times and calculate the RMSE of the posterior mean and variance. 
In the calculation of RMSE the true posterior mean and variance are obtained by using the standard PF with $5\times10^5$ particles.
We show the results in several figures. First, in Fig.~\ref{f:car_alpha} (left) we show 
the weight parameter $a$ computed in one trial
for both DMPF and EnKPF.
We observe rather mixed results in DMPF: the weight $a$ is estimated to be around 1 in about 40 steps and significantly less than 1 in the rest. 
{\color{black} To further understand the behavior of the DMPF method, we consider the time step at $t=72$ where $a$ is estimated to be $0$,
which indicates that the algorithm detects strong non-Gaussianity in the step.
We show the scatter plots of the particles (the particles
are resampled so they all have the same weights) at this step, and 
one can see from the plots that the posterior samples are distributed very differently from Gaussian.} 

We then plot the RMSE of the mean and the variance in Fig.~\ref{f:car_rmse} for all methods,
and we summarize the RMSE averaged over all time steps in Table~\ref{tab:car}. 
{\color{black}The figure shows that the EnKF works well for time steps before $t=70$, and 
its results become significantly inaccurate shortly after that,
which makes its average RMSE over all time steps the worst among all methods.
From Table~\ref{tab:car}, we can see that DMPF and EnKPF have the best performance
among the five methods where the EnKPF yields lower variance RMSE. 
It should be noted here that, a limitation of the EnKPF is that,
its performance depends critically on the constrained diversity $\tau$, which has to be specified by the user.
On the other hand, the DMPF method does not have such free parameters that need to be tuned.} 

\begin{table}[htbp]	
	\caption{RMSE averaged over all time steps in the car-like robot model.}
	\label{tab:car}
	\centering
	\begin{tabular}{|c|c|c|} \hline
		- & \bf mean & \bf variance \\ \hline
		\bf PF  & 0.069 & 0.017 \\ 
		\bf EnKF  & 0.15 & 0.052\\ 
		\bf DMPF & 0.061 & 0.019 \\ 
		\bf EnKPF & 0.059 & 0.013\\  
		\bf WEnKF & 0.066 & 0.017\\ \hline
	\end{tabular}	
\end{table}

\begin{figure}
\centerline{\includegraphics[width=.5\textwidth]{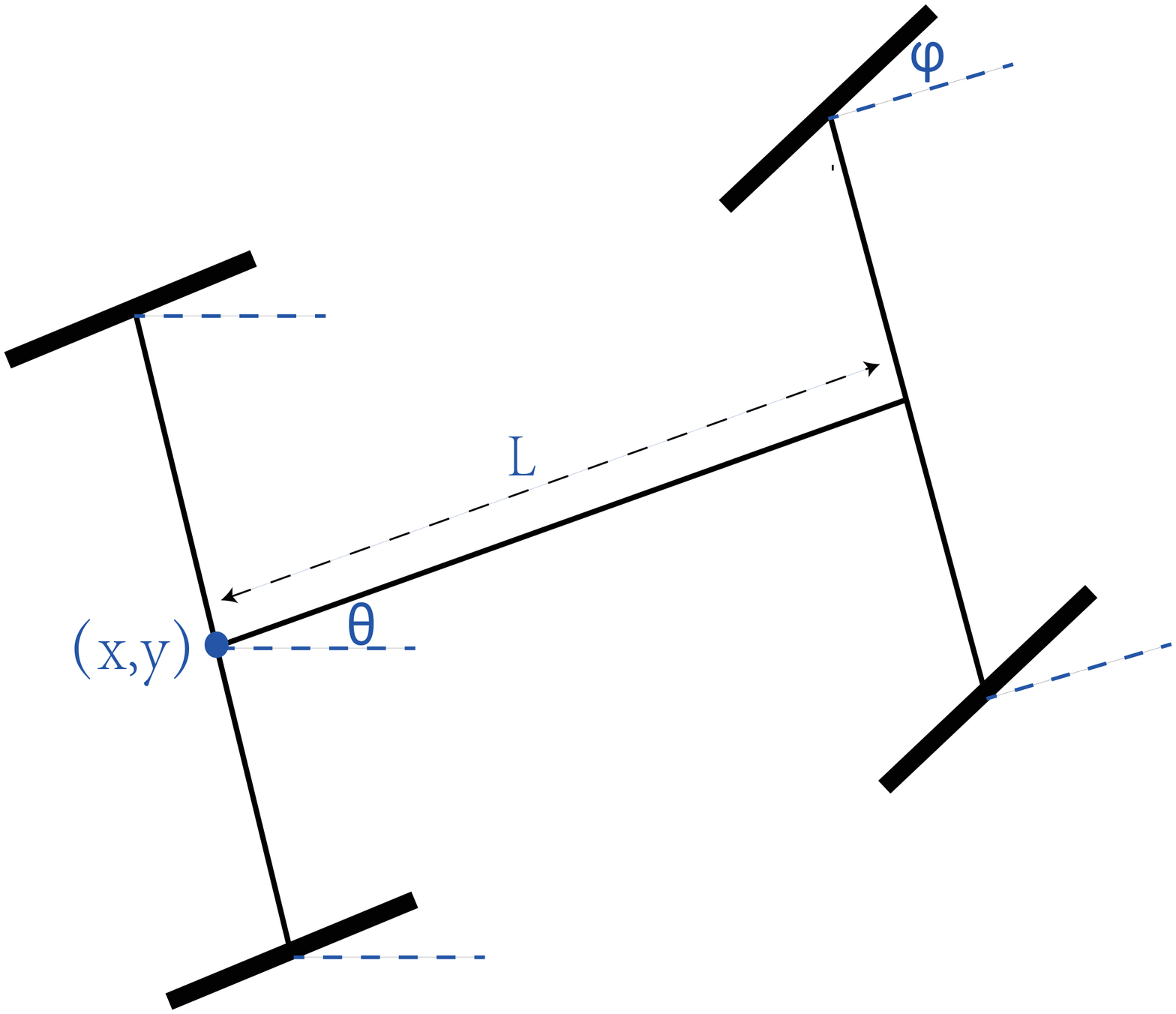}} 
{\color{black}\caption{The schematic illustration of the car-like robot model.}\label{f:robot}}
\end{figure}	

\begin{figure}
\centerline{\includegraphics[width=.95\textwidth]{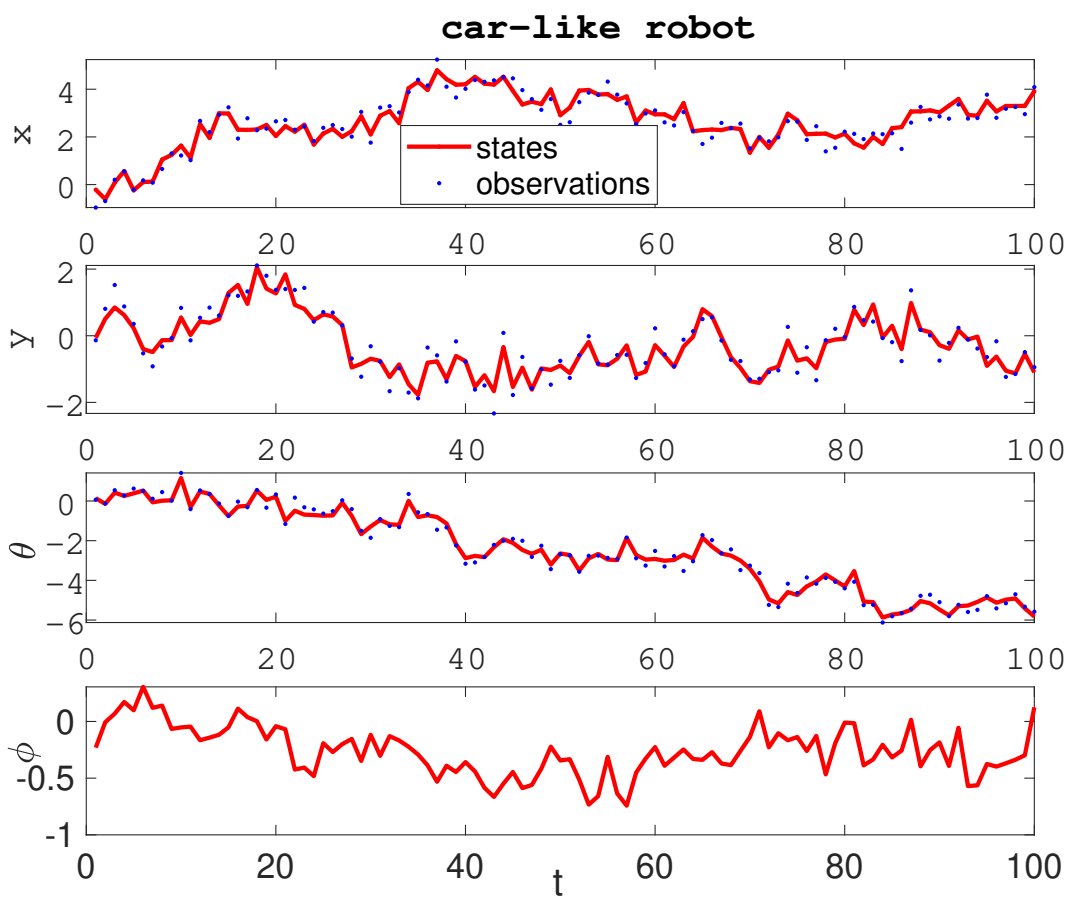}} 
\caption{The true state (dashed lines) and the simulated observations (dots) of the car-like robot model.}\label{f:data_car}
\end{figure}	

\begin{figure}
\centerline{\includegraphics[width=.5\textwidth]{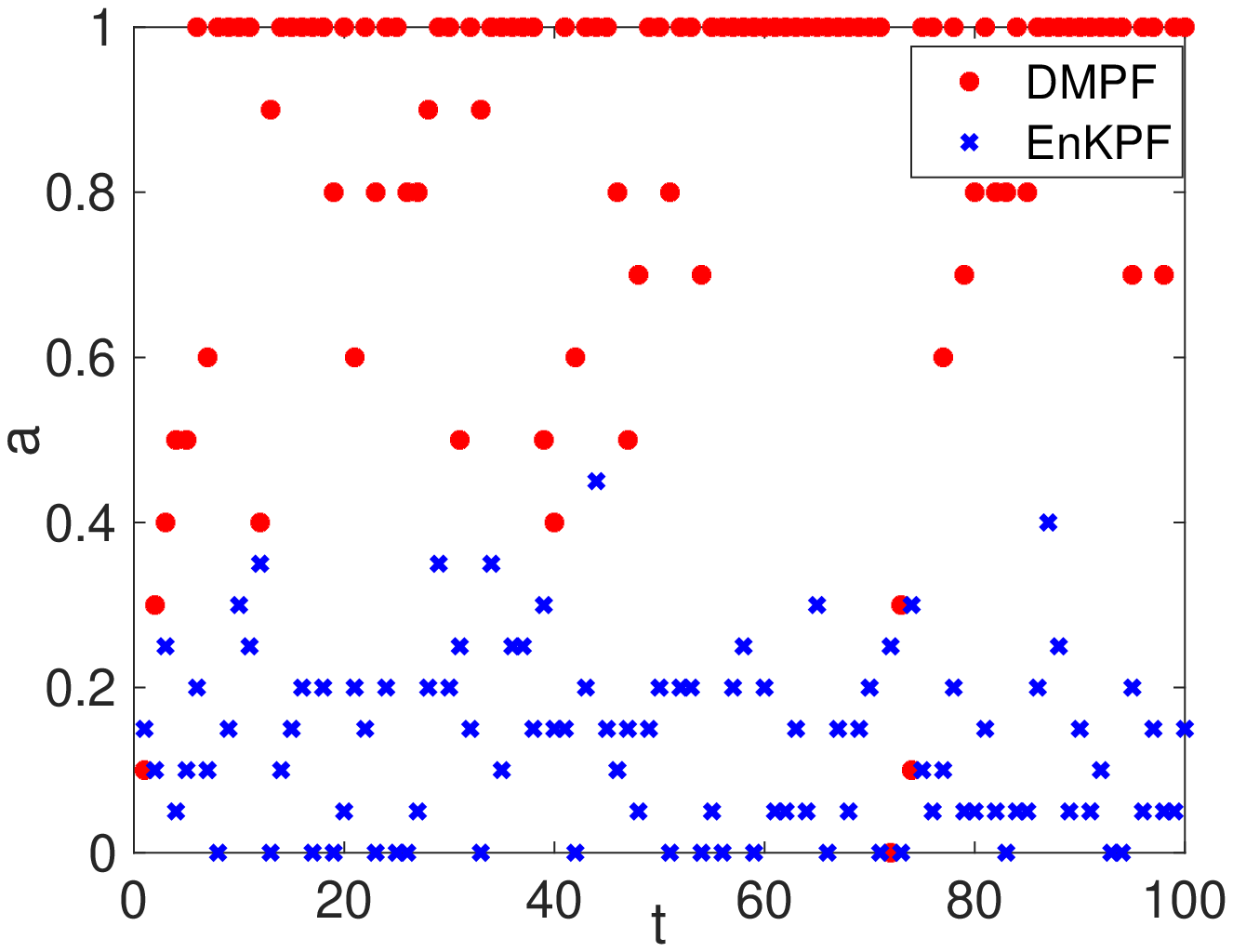}
\includegraphics[width=.5\textwidth]{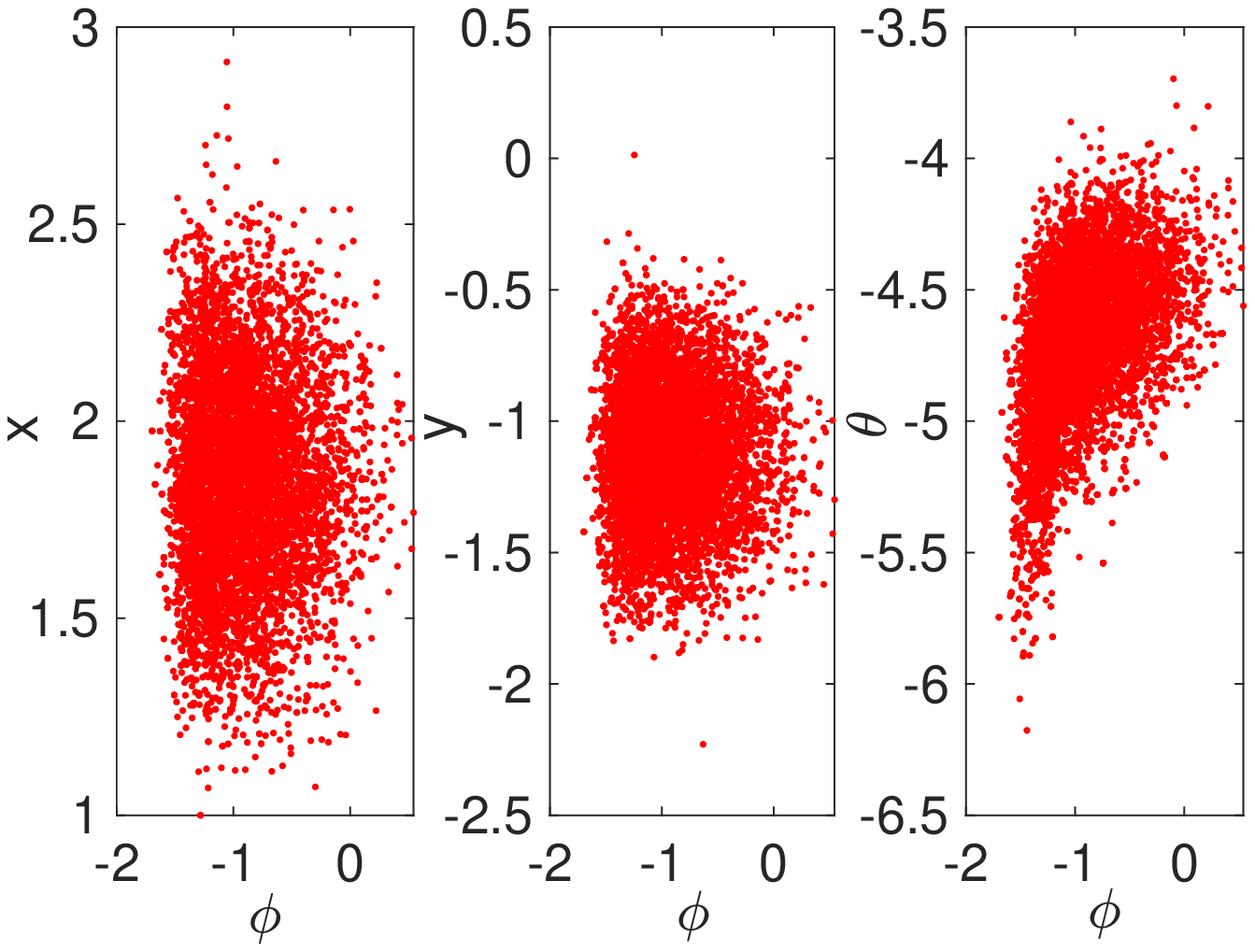}} 
\caption{{\color{black} Left: the parameter $a$ in DMPF and EnKPF for the car-like robot model.}
The scatter plots of the particles at $t=72$ in the DMPF method.}\label{f:car_alpha}
\end{figure}


\begin{figure}
\centerline{\includegraphics[width=.55\textwidth]{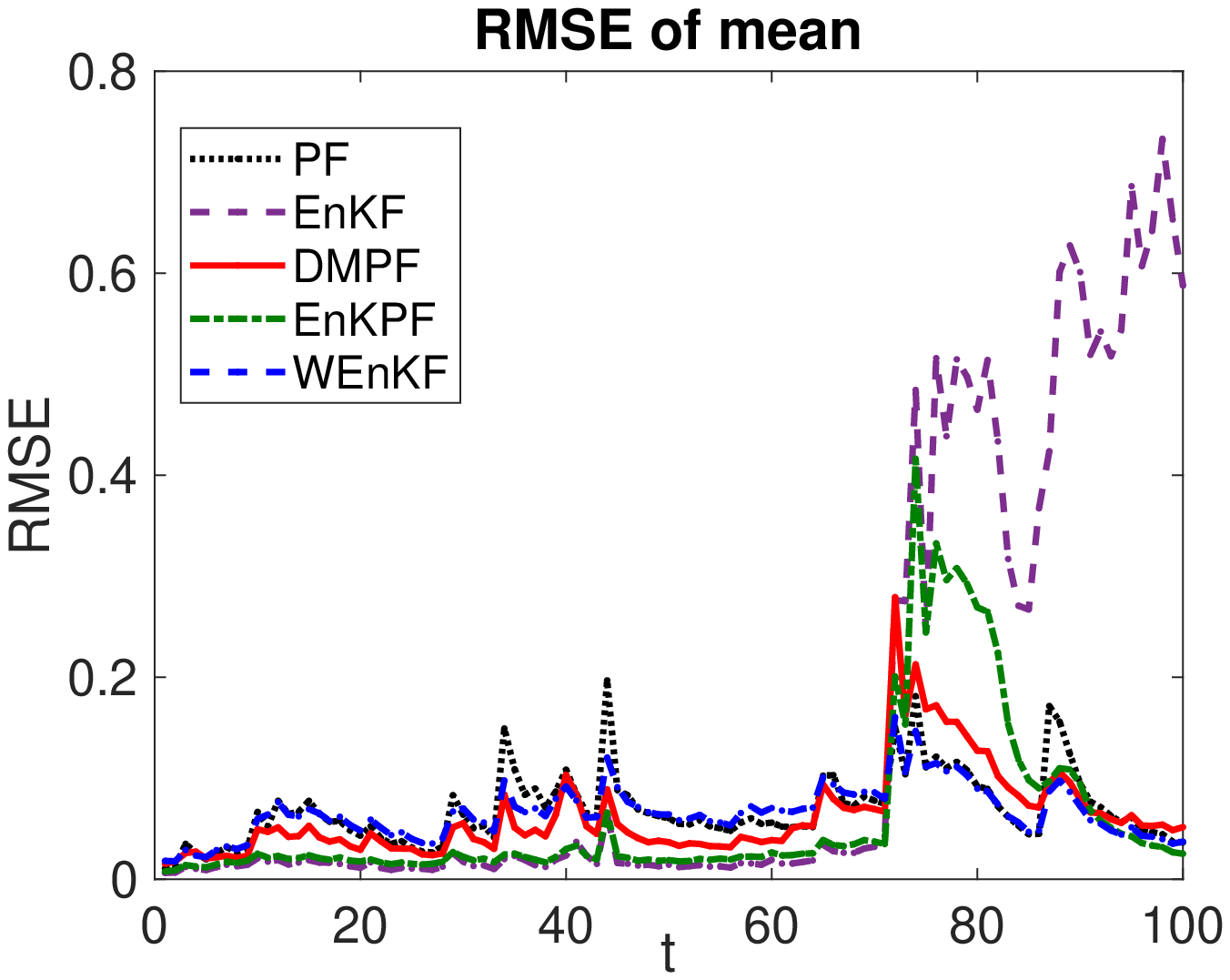}
\includegraphics[width=.55\textwidth]{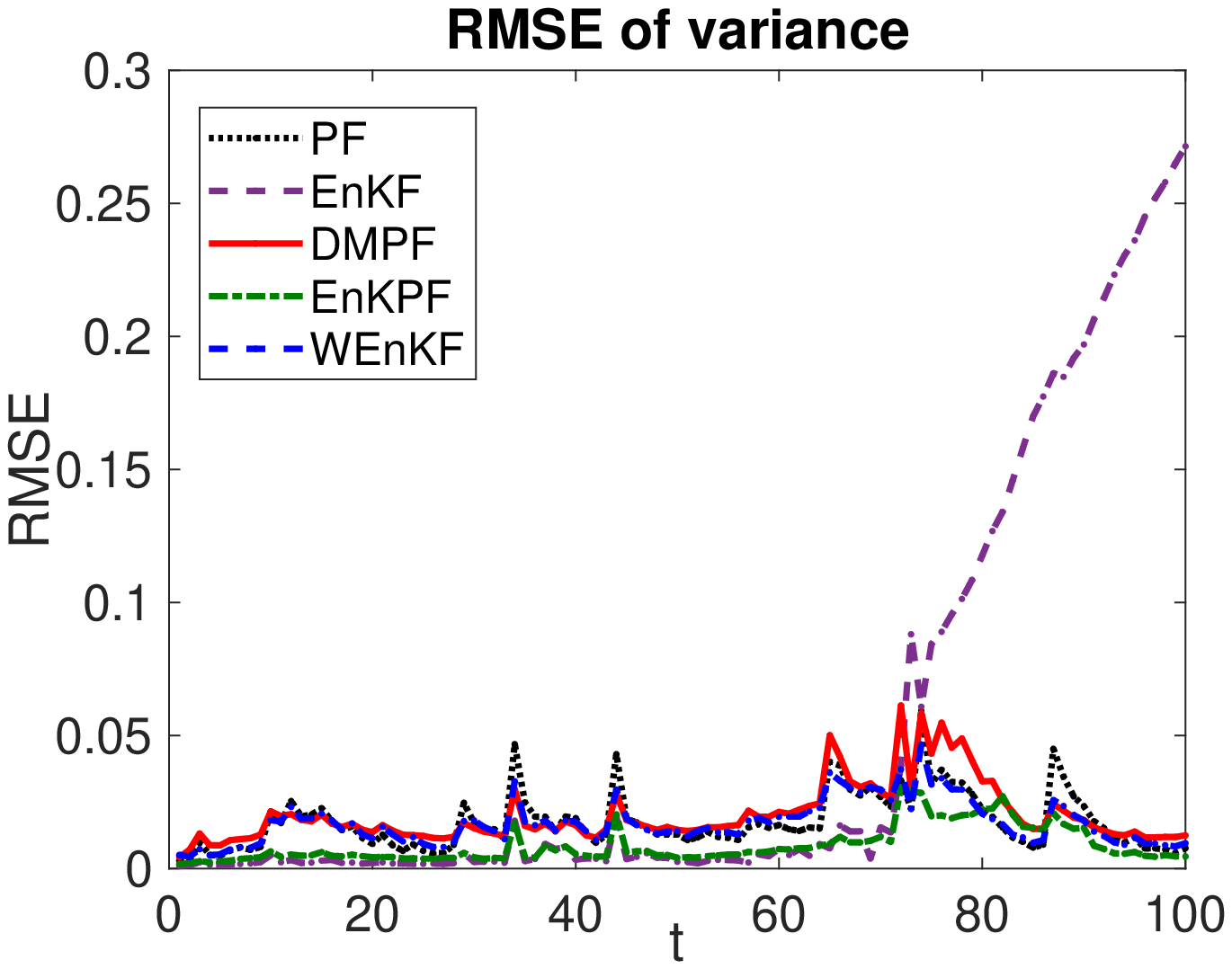}} 
\caption{{\color{black}The RMSE of the posterior mean (left) and of the posterior variance (right) for the car-like robot model.}}\label{f:car_rmse}
\end{figure}
\section{Conclusions} \label{Sec:conclusion}
In summary, we have presented a marginal particle filtering method that samples the posterior distribution in the marginal state space.
 In particular, we propose a defensive scheme to construct a proposal distribution in the marginal space by combining 
 the standard PF and the EnKF-based proposals, which ensures that the algorithm performs well even when the posterior is strongly non-Gaussian.  
 The proposed method can automatically adjust the relative weight of the PF and the EnKF components.
 We provide three examples to demonstrate the performance of the proposed method: in the first example we show that 
 the DMPF method performs well when the standard EnKF fails due to the strong non-Gaussianity of the posterior distribution;
 the second example demonstrates that, the DMPF method can significantly outperform PF when the posterior is close to Gaussian;
finally it was illustrated by the third example that the defensive scheme used in our method can provide better results
than simply use the EnKF as the IS distribution, when a significant fraction of the time steps can not be well approximated by Gaussian. The three examples demonstrate that the DMPF method has 
a good performance regardless of whether the posteriors are close to Gaussian. 
We believe that the method can be useful in a wide range of practical data assimilation problems. 

The proposed method can be improved in several aspects. 
First, in this work we have mainly considered problems where the marginal state space is of rather low dimensions. 
On the other hand, for problems of high dimensions, 
it becomes  challenging to accurately estimate the IS weights in each step. 
This issue needs to be addressed so that the MPF type of methods can apply to high dimensional problems. 
Second, as has been discussed in \cite{klaas2005toward}, computing the IS weights in each time step is of $M^2$ complexity where $M$ is the number of particles,
and as a result the method become prohibitively expensive for problems requiring a large number of particles.   
It has been suggested in  \cite{klaas2005toward} that some approximation techniques such as the fast multipole method~\cite{greengard1987fast} can be used to reduce the computational cost, but further improvement of the efficiency is still needed to make the method useful in large scale problems. 
Finally we reinstate it here that the proposed DMPF scheme does not require the IS distribution to be the EnKF approximation,
 and rather it can be used with any desired IS distribution. 
For example, one can design mixtures to approximate the marginal posteriors~\cite{bengtsson2003toward,stordal2011bridging,chen2000mixture}
and use them as the IS distribution in the defensive scheme. 
Finally, a difficulty in the particle based methods is that the underlying dynamical system is often computationally intensive,
and considerable efforts have been devoted to accelerating the computation, including surrogate models~\cite{li2009generalized,li2014adaptive}, multi-level methods~\cite{jasra2017multilevel}, and dimension reduction techniques~\cite{solonen2016dimension}, just to name a few.
We expect that these techniques can be used to accelerate the DMPF algorithm as well. 
We hope to study these issues and improve the DMPF method in the future. 


\bibliographystyle{siamplain}
\bibliography{mpf}

\end{document}